\documentclass[aps,prl,twocolumn]{revtex4-1}

\usepackage{amssymb} 
\usepackage{amsmath} 
\usepackage{graphicx}
\usepackage{color}
\usepackage{gensymb}
\usepackage{todonotes}
\usepackage{float} 
\usepackage{xcolor} 
\usepackage{cancel}
\usepackage{comment}

\usepackage{animacros}

\pdfoutput=1

\newcommand{\eq}{\begin{equation}}
\newcommand{\eeq}{\end{equation}}
\renewcommand{\figurename}{\textbf{Figure}}

\newcommand{\blue}[1]{{\color{black}#1}}
\newcommand{\red}[1]{{\color{black}#1}}

\begin{document}


\title{Quantum Approximate Optimization \blue{of the Long-Range \\ Ising Model} with a Trapped-Ion Quantum Simulator}

\author {G. Pagano$^{1,2}$, A. Bapat$^{1}$, P. Becker$^{1}$, K. S. Collins$^{1}$,  A. De$^{1}$, P. W. Hess$^{1,3}$, H. B. Kaplan$^{1}$, A. Kyprianidis$^{1}$,W. L. Tan$^{1}$, C. Baldwin$^{1}$, L. T. Brady$^{1}$, A. Deshpande$^{1}$, F. Liu$^{1}$, S. Jordan$^{4}$, A. V. Gorshkov$^{1}$, and C. Monroe$^{1}$}

\affiliation{$^1$ Joint Quantum Institute, Joint Center for Quantum Information and Computer Science, and Physics Department, University of Maryland and National Institute of Standards and Technology, College Park, MD 20742}
\affiliation{$^2$ Department of Physics and Astronomy, Rice University, Houston, TX 77005-1892}
\affiliation{$^3$ Middlebury College Department of Physics, Middlebury, VT 05753, USA}
\affiliation{$^4$ University of Maryland Institute for Advanced Computer Studies, College Park, Maryland 20742, USA}


\begin{abstract}
Quantum computers and simulators may offer significant advantages over their classical counterparts, providing insights into
quantum many-body systems \cite{Feynman1982} and possibly \blue{improving performance for solving} exponentially hard problems \cite{Nielsen2011}, such as optimization \blue{\cite{Farhi2000,Farhi2014}} and satisfiability \cite{Farhi2001}. 
Here we report the implementation of a \blue{low}-depth Quantum Approximate Optimization Algorithm (QAOA) \cite{Farhi2014} using an analog quantum simulator. We estimate the ground state energy of the Transverse Field Ising Model with long-range interactions with tunable range \blue{and we optimize the corresponding combinatorial classical problem by sampling the QAOA output with high-fidelity, single-shot individual qubit measurements. We execute the algorithm  with both an exhaustive search and closed-loop optimization of the variational parameters, approximating the ground state energy with up to 40 trapped-ion qubits. We benchmark the experiment with bootstrapping heuristic methods scaling polynomially with the system size. We observe, in agreement with numerics, that the QAOA performance does not degrade significantly as we scale up the system size, \red{and that the runtime is approximately independent from the number of qubits}. We finally give a comprehensive analysis of the errors occurring in our system, a crucial step in the path forward towards the application of the QAOA to more general problem instances.}
\end{abstract}

\maketitle

\blue{A promising near-term application of quantum devices is the production of highly entangled states with metrological advantage or with properties of interest for many-body physics and quantum information processing.} One possible approach \blue{to produce useful quantum states} is to use quantum devices to perform \blue{adiabatic quantum computing} \cite{Kadowaki1998,Farhi2000}, which in some cases may provide an advantage over classical approaches \blue{\cite{Hastings2013}}. However, \blue{adiabatic quantum computing} has stringent adiabaticity requirements that hinder its applicability on existing quantum platforms that have finite coherence times \cite{Richerme2013}. 

Alternatively, hybrid quantum-classical variational algorithms may approximately solve hard problems in realms such as quantum magnetism, quantum chemistry \cite{mcclean2016}, and high-energy physics \cite{Kokail2018}. This is because the key resource of quantum computers and simulators is quantum entanglement, which is exactly what makes these many-body quantum problems hard. In a hybrid variational algorithm, entangled states are functions of variational parameters that are  iteratively optimized by a classical algorithm. One example is the Quantum Approximate Optimization Algorithm \cite{Farhi2014}, which consists of a ``bang-bang" protocol that \blue{can} provide approximate answers in a time-efficient way, using devices with finite coherence times and without the use of error-correction \cite{harrow2016qaoa,Lloyd2018,Preskill2018,Zhou2018, Hastings2019}. 

Similarly to \blue{adiabatic quantum computing}, the QAOA protocol encodes the objective function of the optimization problem in a target spin Hamiltonian. The optimization steps of the QAOA are based on unitary evolution under the target Hamiltonian and a non commuting ``mixing'' operator. 
\blue{In general,} the QAOA relies on a classical outer loop to optimize the quantum circuit, aided by physical intuition \cite{bapat2018,verdon2018,Jiang2017,Wang2018} or observed structure of the variational parameters \cite{Zhou2018, brady2019,crooks2018,Mbeng2019}, producing fast, low-depth circuits for approximate solutions.
The QAOA has also been proposed as an efficient way to produce entangled quantum states, such as the ground states of critical Hamiltonians, which gives access to their corresponding energies \cite{Ho2018a,Ho2018b}.

In this work, we employ a collection of interacting trapped-ion qubits to experimentally implement a specific instance of the QAOA\blue{, which is native to our quantum hardware}. \blue{We focus on both the energy minimization of the quantum Hamiltonian and the combinatorial optimization of the corresponding classical problem. Both problems are} encoded in the transverse field anti-ferromagnetic Ising Hamiltonian with long-range interactions:
\begin{equation}
H = \underbrace{\sum_{i<j} J_{ij} \sigma^x_i \sigma^x_{j}}_{H_A} + \underbrace{B\sum_i\sigma_i^y}_{H_B}.
\label{eq_1}
\end{equation}
Here we set the reduced Planck's constant $\hbar=1$, $\sigma_i^\gamma$ ($\gamma=x,y,z$) is the Pauli matrix acting on the $i^\text{th}$ spin along the $\gamma$ direction of the Bloch sphere, $J_{ij}>0$ is the Ising coupling between spins $i$ and $j$\blue{, which, in our case, falls off as a power law in the distance between the spins}, and $B$ denotes the transverse magnetic field. It is well-known \cite{Koffel2012} that the Hamiltonian (\ref{eq_1}) exhibits a quantum phase transition for anti-ferromagnetic interactions with power law decay. \blue{One of the goals of this work is to find an approximation of the ground state energy both at the critical point $(B/J_0)_c$, where $J_0$ is the average nearest-neighbour coupling, and in the case of $B=0$, optimizing the QAOA output for the classical Hamiltonian $H_A$.}
The realization of the QAOA entails a series of unitary quantum evolutions (see Fig.~1) under the non-commuting  Hamiltonians $H_A$ and $H_B$ \blue{(defined under Eq.~(\ref{eq_1}))} that are applied to a known initial state $\ket{\psi_0}$. The state obtained after $p$ layers of the QAOA is:
\begin{equation}
\label{eq_}
|\vec{\beta},\vec{\gamma}\rangle=  \prod_{k=1}^{p}e^{-i\beta_k (H_B/J_0)} e^{-i\gamma_k (H_A/J_0)} \ket{\psi_0},
\end{equation}
where the \blue{evolution times (or, henceforth, ``angles")} $\beta_k$ and $\gamma_k$ are variational parameters used in the $k$-th QAOA layer to minimize the final energy $E(\vec{\beta},\vec{\gamma})=\langle \vec{\beta}, \vec{\gamma} |H|  \vec{\beta}, \vec{\gamma}\rangle$.

\begin{figure}[t!]
\centering
\includegraphics[width=1\columnwidth,angle=-0]{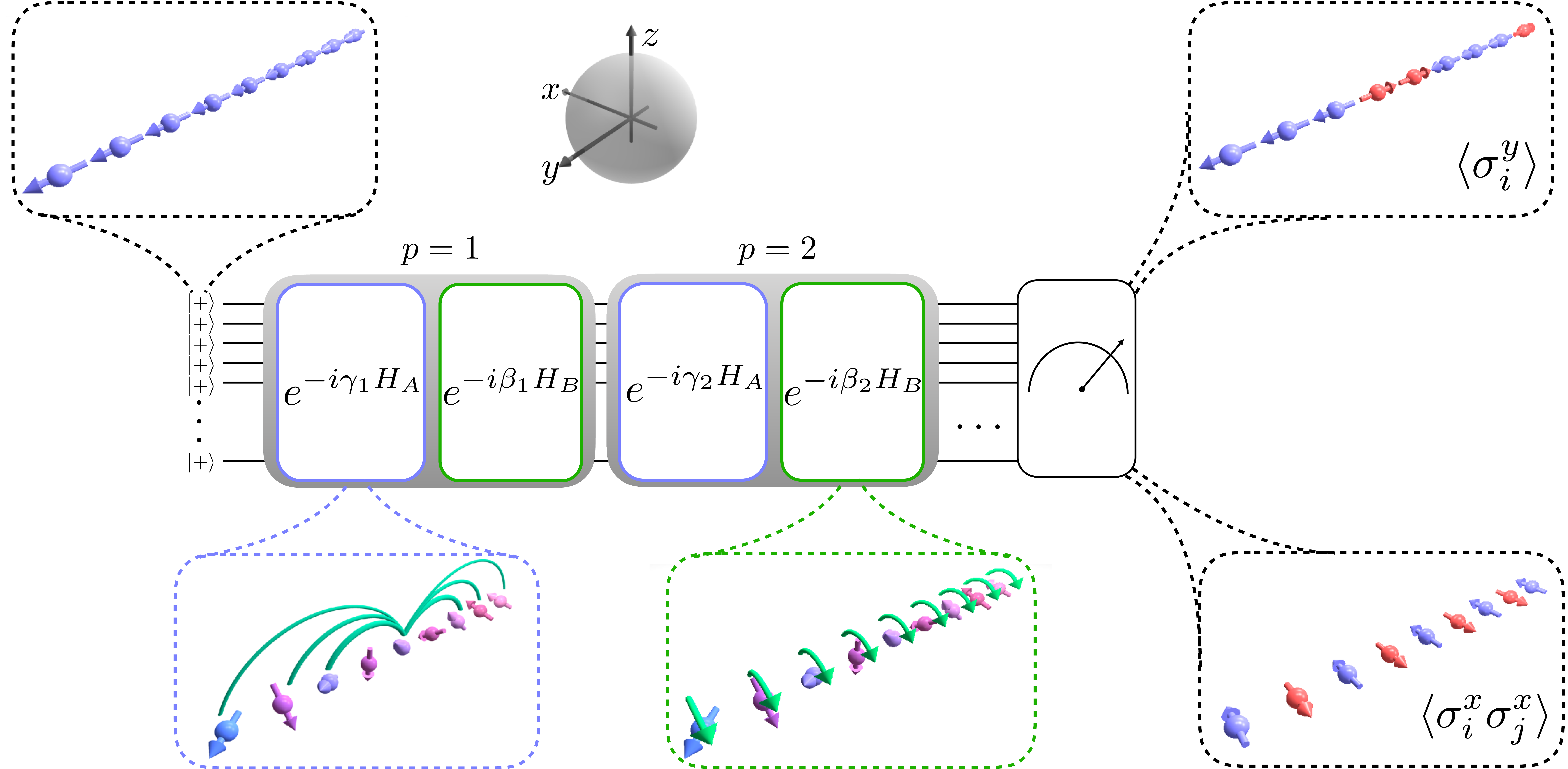}
\caption{\textbf{QAOA protocol. } The system is initialized along the $y$ direction in the Bloch sphere in the $\ket{+}^{\otimes N}$ state. The unitary evolution under $H_{A(B)}$ is implemented for angles $\gamma_i (\beta_i$) for $p$ times. At the end of the algorithm global measurements in the $x$ and the $y$ basis are performed to compute the average energy $\langle H \rangle=E(\vec{\beta},\vec{\gamma})$, which is compared to the theoretical ground state energy $E_{gs}$. }
\label{fig1}
\end{figure}

In order to implement the quantum optimization algorithm, each spin in the chain is encoded in the $^2$S$_{1/2}$ $\ket{F=0,m_F=0} \equiv |\!\!\downarrow\rangle_z$ and $\ket{F=1,m_F=0} \equiv |\!\!\uparrow\rangle_z$ hyperfine ``clock" states of a $^{171}$Yb$^+$ ion 
(see Supplementary). In this work, depending on the number of qubits and measurements required, we employ two different quantum simulation apparatus to run the QAOA, which will herein be referred to as system 1 ~\cite{Kim2009} and system 2 ~\cite{Pagano2019} (see Supplementary). Both systems are based on a linear rf Paul trap where we store chains of up to $N=40$ ions  and initialize the qubits in the ground state of $H_B$, namely the product state $\ket{\!\! \uparrow \uparrow \cdots \uparrow}_y\equiv\ket{+}^{\otimes N}=\ket{\psi_0}$, where $\ket{\!\!\uparrow}_y \equiv (\ket{\!\!\uparrow}_z+ i \ket{\!\!\downarrow}_z)/\sqrt{2}$ and $B$ is assumed to be negative. The unitary evolution under $H_A$ is realized by generating spin-spin interactions through spin-dependent optical dipole forces implemented by an applied laser field. This gives rise to effective long-range Ising couplings that fall off approximately as $J_{ij} \approx J_\textrm{0}/|i-j|^\alpha$~\cite{Porras2004}. The power-law exponent $\alpha \sim 1$ and the interaction strengths vary in the range $J_0/2\pi = $(0.3-0.57) kHz, depending on the system size and the experimental realization (see Supplementary for details). The unitary evolution under $H_B$ is generated by applying a global  rotation around the $y$-axis of the Bloch sphere.

After each run of the algorithm, we perform a projective measurement of each spin in the $x\,(y)$ basis to measure $\langle H_A\rangle$ ($\langle H_B\rangle$) (see Fig.~\ref{fig1}). Measurements in the $x$ and $y$ bases are carried out by performing a $\pi/2$ rotation about the $y$($x$)-axis of the Bloch sphere, illuminating the ions with resonant laser light, and collecting the $\sigma_i^z$-dependent fluorescence on a camera with site-resolved imaging. The energy is calculated by combining the measurements of the two-body correlators $\langle \sigma_i^x \sigma_j^x\rangle$ and the total magnetization along the $y$ axis $\sum_i \langle \sigma^y_i \rangle $, where the indices $i,j$ range from 1 to $N$. We benchmark the experimental outcome $E(\vec{\beta},\vec{\gamma})$ with the ground state $E_{gs}$ of the target Hamiltonian (see Eq. \ref{eq_1}) calculated numerically with exact diagonalization or Density Matrix Renormalization Group (DMRG) \cite{Jaschke2018}. In order to quantify the performance of the QAOA, we use the dimensionless quantity 
\begin{equation}
\eta\equiv \frac{E(\vec{\beta},\vec{\gamma})-E_{max}}{E_{gs}-E_{max}},
\label{eq_eta}
\end{equation}
where $E_{max}$ is the energy of the highest excited state. This choice maps the entire many-body spectrum to the $[0,1]$ interval. In the following we show that the best experimental performance $\eta^*$ is close to the theoretical performance $\eta_{th}$, which itself is less than unity for a finite number $p$ of QAOA layers.
\renewcommand{\thefigure}{2}
\begin{figure*}[t!]
\centering
\includegraphics[width=2\columnwidth]{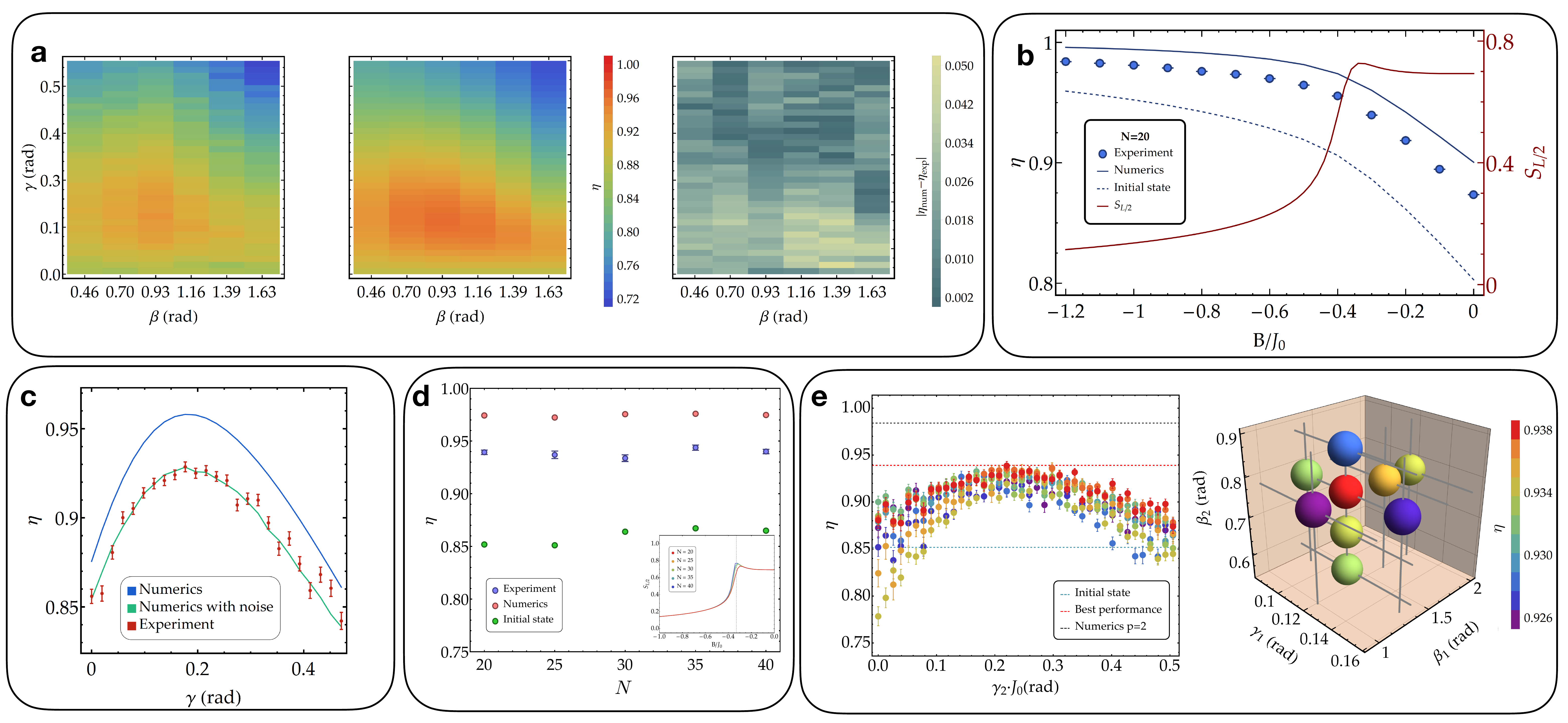}
\caption{\textbf{Exhaustive search for optimal performance.} {\bf (a)} Experimental (left) and theoretical \blue{(center)} performance landscape \blue{and their absolute difference (right)} as a function of the variational parameters $\beta$ and $\gamma$ for $N=20$ qubits ($J_0/2\pi = 0.57$ kHz, $B/J_0\sim-0.3$) \blue{, displaying an average absolute difference of 1.9\% over 210 different $\{\beta,\gamma\}$ pairs}. The optimal performance is $\eta^* = (93.8\pm0.4)\%$, whereas the theoretical performance is $\eta_{th}=96.1\%$. Each data point is the result of 1100 (800) experimental repetitions to measure in the $x \,(y)$ basis (data taken on system 1). {\bf (b)} Exhaustive search optimization as a function of $B/J_0$ (see Eq. (1)) (data taken on system 1). The dark red solid line is the half-chain entanglement entropy $S_{L/2}$ computed numerically with DMRG. The dashed blue line represents the performance of the initial product state $\ket{\psi_0}$. 
\blue{{\bf (c)} Comparison between experimental performances and numerics for $B/J_0\sim-0.3$ and $N=12$ as a function of $\gamma$ and $\beta^*=1.12$. Taking into account bit-flip errors and slow drifts in the experimental parameters explains well the discrepancy between experimental and ideal performance (see Supplementary for details).}
{\bf (d)} The $p=1$ QAOA performance as a function of system size $N$ up to 40 qubits (data taken on system 2). Comparison between QAOA experimental and theoretical performance for $B/J_0 \sim -0.3$. Green points show the baseline performance of the initial state $\ket{\psi_0}$. Inset: Convergence of the entanglement entropy peak as a function of number of qubits (see Supplementary). {\bf (e)} $p=2$ exhaustive search for $N=20$ and $B/J_0\sim-0.3$. Left: every color corresponds to a fine scan of $\gamma_2$ with a different set of variational parameters $\beta_1,\beta_2$ and $\gamma_1$ (data taken on system 2). Right: 3D color plot of the performance $\eta$, optimized over $\gamma_2$, as a function of the parameters $\beta_1,\beta_2$ and $\gamma_1$. The best outcome is $\eta^* = (93.9\pm 0.3)\%$ (colored red), whereas the theoretical performance is $\eta_{th}=98.4\%$ (see main text for details). \blue{In {\bf (b),(c),(d)} and {\bf(e)} the error bars are calculated by using the standard deviation from the mean of the measured performance.}
}
\label{fig2}
\end{figure*}

\blue{\textit{Quantum Hamiltonian Optimization -} We first focus on the $p=1$ optimization of the full quantum problem}, where two variational parameters ($\gamma$ and $\beta$) are used to minimize the energy \blue{of the Hamiltonian (1)}. In this case, the time-evolved one- and two-point correlation functions can be efficiently computed \cite{dylewsky2016, Hadfield2018}. This leads to a general formula for the energy expectation under a state produced by the $p=1$ QAOA that is used to compute the theoretical performance of the algorithm (see Supplementary). In Fig. \ref{fig2}a we show an experimental exhaustive search over the parameter space $\{\gamma,\beta\}$ and compare it to the theoretical performance of the algorithm, showing good agreement for $N=20$ qubits. We also compare the performance of our algorithm as a function of $B/J_0$ with the expected QAOA performance $\eta_{th}$ (see Fig. \ref{fig2}b). 

\renewcommand{\thefigure}{3}
\begin{figure*}[t]
\centering
\includegraphics[width=2.05\columnwidth]{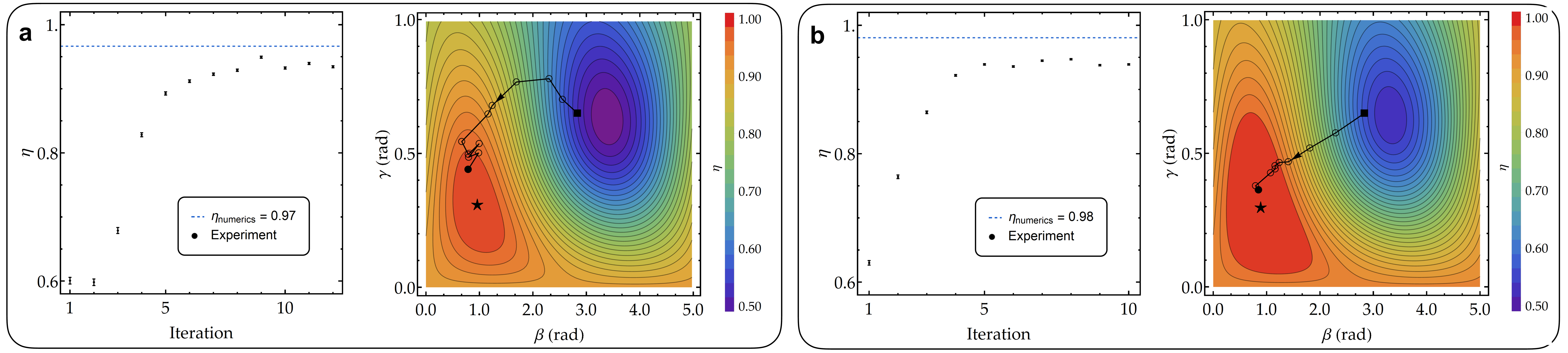}
\caption{\textbf{Gradient descent search for p=1 QAOA.} (a) $N=12$ and (b) $N=20$. Left: performance $\eta$ convergence as a function of iterations of the classical-quantum hybrid algorithm with (a) $N=12$ ($ J_0/2\pi = 0.57$ kHz, $B/J_0=-0.3$) with a measured $\eta^*=(94.9\pm 0.2) \% $ and (b) $N=20$ qubits ($J_0/2\pi = 0.55$ kHz, $B/J_0=-0.3$)  with a measured $\eta^*=(94.7\pm 0.1)\%$. Right: the algorithm trajectory on the theoretical performance landscape plotted as a function of $\gamma$ and $\beta$. Each energy evaluation takes 4000 (6000) shots for 12 (20) qubits. The error bars are standard deviation from the mean of the measured performance (data taken on system 1).}
\label{fig3}
\end{figure*}

As shown in Ref. \cite{Koffel2012}, for transverse field greater than the critical value, the ground state is a low entanglement paramagnet, whereas below the critical point the ground state is an entangled superposition of anti-ferromagnetic states.  We locate this critical point at $|B/J_0|=0.31$ for 20 qubits by computing the half-chain entanglement entropy $S_{L/2}=-\mathrm{Tr}(\rho_{L/2}\log{\rho_{L/2}})$ of the ground state numerically, where $\rho_{L/2}$ is the half-chain reduced density matrix. As shown in Fig. \ref{fig2}b, while the experimental performance is $\eta>94\%$ when $|B/J_0|$ is above the critical point, the gain relative to the initial state $\ket{\psi_0}$ is modest. On the other hand, below the critical point, the target state is more entangled, which allows for a larger experimental performance gain, at the expense of a reduced absolute performance. \blue{In order to quantitatively assess the gain over the finite initial state performance, we introduce a performance natural scale based on the quantity $\sigma_\eta(J_0,B,N)$, namely the standard deviation around the mean performance achieved implementing a QAOA algorithm with random angles (see Supplementary for details). For $N=20$ and $B/J_0\sim-0.3$, $\sigma_{\eta}\sim 2\times10^{-3}$. Our experimental performance at the critical point $\eta^*$ is more than 20$\sigma_\eta$ away from the initial state. On the other hand, the discrepancy between the ideal and experimental performance can be explained by taking into account our noise sources in the numerics (see Fig. 2c and the Combinatorial Optimization section below).}

We investigate the performance of the $p=1$ QAOA algorithm as a function of the number of qubits. For each system size, we ensure that the spin-spin couplings $J_{ij}$ have the same dependence on the qubit distance $|i-j|$ by varying the trap parameters (see Supplementary). \blue{As shown in the inset of Fig. \ref{fig2}d}, the half-chain entanglement entropy as a function of system size $N$ exhibits a peak located at $B/J_0\sim-0.33$, displaying the onset of the phase transition as $N$ tends to infinity. For all system sizes, we optimize the algorithm by performing a scan of the interaction angle $\gamma$ and applying discrete variations of the mixing angle $\beta$ around the optimal value predicted by the theory. In Fig.~\ref{fig2}d we compare the optimal experimental and theoretical performances $\eta$ for different system sizes from 20 up to 40 qubits for fixed $B/J_0 \sim -0.3$. \red{We observe experimentally that the QAOA yields a similar performance as a function of number of qubits even if the algorithm runtime stays approximately constant as the number of qubits increases. Numerically, we found that the performance $\eta$ scales polynomially with $N$ and with the number of layers $p$ (see Supplementary). \blue{Assuming extrapolation to higher numbers of qubits holds,} this scaling, combined with a polynomial-time search heuristic, suggests that for any desired energy threshold $\epsilon$, our approach allows us to approximate the energy to a degree $\eta > 1-\epsilon$ in time and number of layers that scale as $\text{poly}(N,1/\epsilon)$.}


We experimentally perform a search for the optimal $p=2$ QAOA performance using 20 qubits. Unlike the $p=1$ case, there is no known analytic formula to efficiently compute the energy. 
However, exploiting relationships between optimal angles as a function of increasing $p$, we use a bootstrapping heuristic (see Supplementary for details) that allows the experiment to identify a set of optimal angles faster than a global parameter search. 
\blue{The bootstrapping heuristic computes a guess for optimal angles at $p$ given optimal angles at lower $p$. A local optimizer, such as the greedy gradient descent described below, is then needed to take this guess to the true optimum. Our new heuristic method allows us to find variational parameters in time that scales polynomially with the number of layers and sublinearly in the number of qubits (when used in conjunction with the quantum device).}

We start from the optimal guess and perform a fine scan of $\gamma_2$, while varying $\gamma_1, \beta_1$ and $\beta_2$ in larger steps. The result is shown in Fig.~\ref{fig2}d, where we plot the performances $\eta$ as a function of $\gamma_2$ for every set of parameters used in the experiment. Fig.~\ref{fig2}d shows also a colour plot of all the optimal energies found as a function of the other three parameters $\gamma_1, \beta_1$ and $\beta_2$. The $p=2$ QAOA performance with 20 qubits $\eta^*=(93.9 \pm 0.3)\%$ is in agreement with the $p=1$ performance in system 2, taken with the same parameters (see Fig.~\ref{fig2}c). This indicates that decoherence and bit-flip errors (see Supplementary) accumulated during longer evolution times are already balancing out the $2\%$ expected performance gain of one additional optimization layer.

As a brute force approach is inefficient, we implement a closed-loop QAOA by interfacing the analog trapped-ion quantum simulator with a greedy gradient-descent algorithm to optimize the measured energy. In the $p=1$ QAOA, we can visualize the optimization trajectory on the theoretical performance surface as shown in Fig. \ref{fig3}. \blue{Starting from a guess $(\beta^{(0)}, \gamma^{(0)})$, we measure the approximate local gradient by performing the energy measurements in two orthogonal directions $\beta^{(0)}+\delta\beta$ and $\gamma^{(0)}+\delta\gamma$ to compute the new guess $(\beta^{(1)},\gamma^{(1)})$, where we measure the new energy on the quantum simulator.} As shown in Fig.~\ref{fig3}, the algorithm converges after about 10 iterations.
Compared to an exhaustive search, the gradient descent uses fewer queries to the quantum simulator and is therefore more robust to slow drifts in the experimental system. For this reason, we are able to achieve a better performance compared to the exhaustive search method.

\renewcommand{\thefigure}{4}
\begin{figure*}[t]
\centering
\includegraphics[width=2.05\columnwidth]{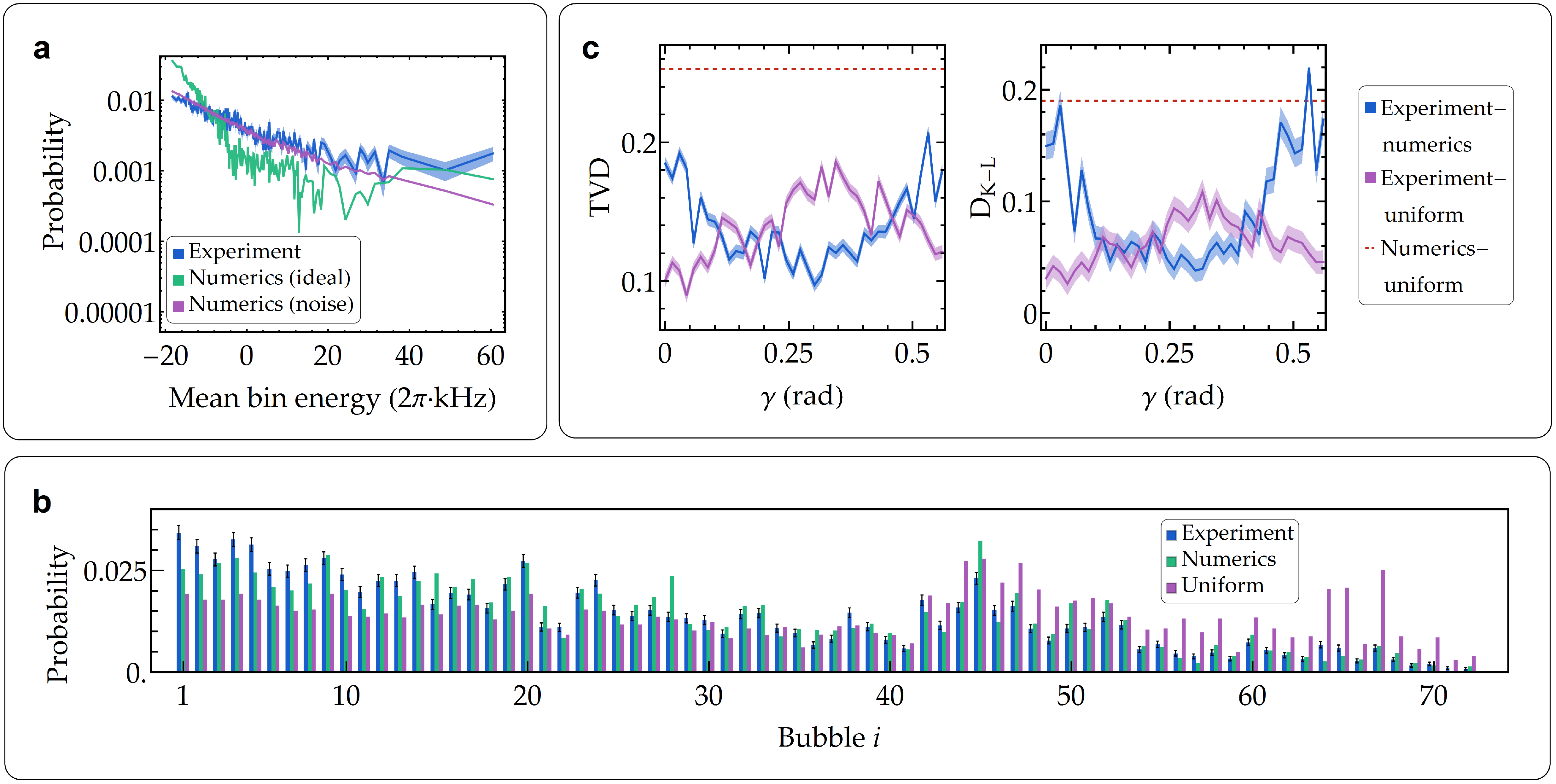}
\caption{\textbf{Sampling from $p=1$ QAOA}. {\bf(a)} Eigenstate probability histogram for 12 qubits with $B=0$. The numerical histogram is computed by decomposing the ideal QAOA output state on the $\{|x_i\rangle\}$ basis. We performed 10800 measurements to oversample the Hilbert space of dimension $2^N=4096$ at the optimal parameters \blue{$\beta^*=0.25$ and $\gamma^*=0.31$}. \blue{The 4096 eigenstates are grouped in bins of 20 for clarity purposes. The uncertainty bands follow the multinomial distribution standard deviation. Here $J_0/2\pi = 0.33$ kHz (see noise sources section in Supplementary for details).} {\bf(b)} Histogram of coarse-grained distributions (see main text for details) comparing data, theory and the uniform distribution. \blue{The error bars here also represent the standard deviation of the multinomial distribution.} {\bf(c)} Total Variation Distance and Kullback-Leibler divergence as a function of $\gamma$, keeping $\beta$ fixed at the optimal value. The distance from the uniform distribution increases as the $\gamma$ parameter reaches the optimal point $\gamma^*$. \blue{Dashed lines are the comparison between the ideal distribution for $\{\beta^*, \gamma^*\}$ and the uniform distribution. The uncertainty bands are based on the aforementioned error in the probability of each state bubble for the experimental distribution, propagated to the TVD and the $\mathrm{D_{K-L}}$ according to Eq. (\ref{eq_TDV})} (data taken on system 2).}
\label{fig4}
\end{figure*}
\blue{\textit{Combinatorial Optimization -}}
\blue{We further explore the performance of the trapped-ion system by investigating the combinatorial optimization of the classical Hamiltonian $H_A$ (see Eq. (1) with $B=0$) approximately sampling the output of the $p=1$ QAOA, using high-fidelity, single-shot measurement of all the qubits.} It has been proven, under reasonable complexity-theoretic assumptions, that no classical algorithm can efficiently sample exactly from a sufficiently general class of $p=1$ QAOA circuits \cite{harrow2016qaoa}. Recent results \cite{Bremner2016, Bouland2019} suggest that this could also hold in the case of approximate sampling (see Supplementary). In this case, by measuring in the $x$ basis, it is possible to sample the probability distribution of all the $2^N$ eigenstates $\ket{x_i}$ of the Hamiltonian $H_A$. We performed the experiment with 12 qubits so that we can both compute the expected QAOA theoretical output and also experimentally over-sample the Hilbert space of all the possible $2^{12}=4096$ possible outcomes. In Fig. \ref{fig4}a we show \blue{on a log scale the QAOA eigenstates probability distribution} using the optimal variational parameters ${\beta^*, \gamma^*}$ and compare the experimental eigenstate histogram with the exact diagonalization prediction of the QAOA output state, \blue{sorting the eigenstates according to their energies}.

However, sampling from the full QAOA output distribution is a daunting task, since the experimental outcome is extremely sensitive to fluctuations in the Hamiltonian parameters and to experimental errors caused by detection and phonon-assisted bit-flip events \blue{and unwanted effective magnetic fields along the $z$ direction of the Bloch sphere caused by uncompensated light shift (see also Supplementary). Given our measured experimental parameters, we can calculate the effect of these errors on the quantum evolution, resulting in a good agreement with the experimental outcome, as shown in Fig. 4a.}

\blue{Another useful way to compare numerics and experimental data is to implement} the coarse-graining procedure of the Hilbert space proposed in Ref. \cite{Wang2015}. After sorting in decreasing order the observed states according to their experimental probability, we iteratively group the states into ``bubbles'' of Hamming distance $L$ around the most probable state, producing a coarse-grained dataset. We then apply the same coarse-graining to the theoretical probability distribution and plot the comparison in Fig. \ref{fig4}b. In this procedure the Hamming distance radius is varied to ensure that each bubble contains a comparable number of experimental shots, leading to bubbles of average Hamming distance $\bar{L}=2.5$. In order to quantitatively compare the coarse-grained experiment and the theory, we \blue{use} two different metrics, namely the total variation distance (TVD) and the Kullback-Leibler divergence ($\mathrm{D_{K\rm{-}L}}$), defined as:
\begin{eqnarray}
\label{eq_TDV}
\rm{TVD}&=&\frac{1}{2}\sum_i | p_i-q_i |, \\
\rm{D_{K\rm{-}L}}&=& - \sum_i p_i \log{\left(\frac{q_i}{p_i}\right)},
\end{eqnarray}
 where $p_i (q_i)$ is the experimental (theoretical) probability of observing the $i$-th outcome.  As shown in Fig.~\ref{fig4}c, when the system is in the initial state, it is closer to a uniform probability distribution since $\ket{\psi_0}$ is an equal superposition of all the eigenstates of $H_A$. On the other hand, as the $\gamma$ parameter is scanned, we observe a net decrease of both TVD and $\mathrm{D_{K-L}}$ \blue{between the experiment and the numerical minimum}, in agreement with the decrease in energy, computed by measuring one and two-body correlators. 

The variational quantum algorithm reported here, with up to 40 trapped-ion qubits, is the largest ever realized on a quantum device. We approximate the ground state energy of a non-trivial quantum Hamiltonian showing almost constant time scaling with the system size. Single-shot high-efficiency qubit measurements in different bases give access to the full distribution of bit-strings that is difficult or potentially impossible to model classically.
With the addition of individual control over the interactions between qubits \blue{as well as improvements to fidelity and system size}, the variational quantum-classical hybrid approach can be employed in this experimental platform to give insight into quantum chemistry \cite{Kandala2017,Hempel2018,Nam2019} and hard optimization problems \cite{Rigetti2017}, such as Max-SAT or exact cover \cite{Farhi2001}, or be used for the production of highly entangled states of metrological interest \cite{Giovannetti2011}.

\subsection{Acknowledgements}
We acknowledge illuminating discussions with L. Duan, Y. Wu, B. Fefferman and S. Wang. The DMRG simulations were performed using Open Source Matrix Product States \cite{Jaschke2018}. This work is supported by the ARO and AFOSR QIS and Atomic and Molecular Physics Programs, the AFOSR MURIs on Quantum Measurement/Verification and Quantum Interactive Protocols, the IARPA LogiQ program, the ARO MURI on Modular Quantum Systems, the ARL Center for Distributed Quantum Information, the NSF QIS program and the NSF Physics Frontier Center at JQI.
A.B., A. Des., F.L., and A.V.G. were mainly supported by the DoE BES QIS program (award No. DE-SC0019449), with additional support by the NSF PFCQC program, DoE ASCR Quantum Testbed Pathfinder program (award No. DE-SC0019040), ARO MURI, AFOSR, NSF PFC at JQI, and ARL CDQI. S. J. and A.B. were supported by the DOE HEP QuantISED Program,  and Quantum Algorithms Teams (QOALAs) programs.

\subsection{Author Contributions}
G.P., A.K., P.B., W.L.T., H.B.K., K.S.C., A.D., P.W.H. and C.M. all contributed to experimental design, construction, data collection and analysis. A.B., L.T.B., F.L., A.D., C.B., S.J. and A.V.G. contributed to the theory for the experiment. All authors contributed to this manuscript.

\subsection{Author Information}
The authors declare competing financial interests:
details are available in the online version of the paper. Readers are welcome to comment on the online version of the paper. Correspondence and requests for materials should be addressed to G.P. (pagano@umd.edu)

\bibliographystyle{naturemag.bst}
\bibliography{QAOA}

\begin{thebibliography}{10}
\expandafter\ifx\csname url\endcsname\relax
  \def\url#1{\texttt{#1}}\fi
\expandafter\ifx\csname urlprefix\endcsname\relax\def\urlprefix{URL }\fi
\providecommand{\bibinfo}[2]{#2}
\providecommand{\eprint}[2][]{\url{#2}}

\bibitem{Feynman1982}
\bibinfo{author}{Feynman, R.~P.}
\newblock \bibinfo{title}{Simulating physics with computers}.
\newblock \emph{\bibinfo{journal}{International Journal of Theoretical
  Physics}} \textbf{\bibinfo{volume}{21}}, \bibinfo{pages}{467--488}
  (\bibinfo{year}{1982}).

\bibitem{Nielsen2011}
\bibinfo{author}{Nielsen, M.~A.} \& \bibinfo{author}{Chuang, I.~L.}
\newblock \emph{\bibinfo{title}{Quantum Computation and Quantum Information:
  10th Anniversary Edition}} (\bibinfo{publisher}{Cambridge University Press},
  \bibinfo{address}{New York, NY, USA}, \bibinfo{year}{2011}),
  \bibinfo{edition}{10th} edn.

\bibitem{Farhi2000}
\bibinfo{author}{{Farhi}, E.}, \bibinfo{author}{{Goldstone}, J.},
  \bibinfo{author}{{Gutmann}, S.} \& \bibinfo{author}{{Sipser}, M.}
\newblock \bibinfo{title}{{Quantum Computation by Adiabatic Evolution}}.
\newblock \emph{\bibinfo{journal}{arXiv e-prints}}
  \bibinfo{pages}{quant--ph/0001106} (\bibinfo{year}{2000}).
\newblock \eprint{quant-ph/0001106}.

\bibitem{Farhi2014}
\bibinfo{author}{{Farhi}, E.}, \bibinfo{author}{{Goldstone}, J.} \&
  \bibinfo{author}{{Gutmann}, S.}
\newblock \bibinfo{title}{{A Quantum Approximate Optimization Algorithm}}.
\newblock \emph{\bibinfo{journal}{arXiv e-prints}}
  \bibinfo{pages}{arXiv:1411.4028} (\bibinfo{year}{2014}).
\newblock \eprint{1411.4028}.

\bibitem{Farhi2001}
\bibinfo{author}{Farhi, E.} \emph{et~al.}
\newblock \bibinfo{title}{A quantum adiabatic evolution algorithm applied to
  random instances of an np-complete problem}.
\newblock \emph{\bibinfo{journal}{Science}} \textbf{\bibinfo{volume}{292}},
  \bibinfo{pages}{472--475} (\bibinfo{year}{2001}).

\bibitem{Kadowaki1998}
\bibinfo{author}{{Kadowaki}, T.} \& \bibinfo{author}{{Nishimori}, H.}
\newblock \bibinfo{title}{{Quantum annealing in the transverse Ising model}}.
\newblock \emph{\bibinfo{journal}{\pre}} \textbf{\bibinfo{volume}{58}},
  \bibinfo{pages}{5355--5363} (\bibinfo{year}{1998}).
\newblock \eprint{cond-mat/9804280}.

\bibitem{Hastings2013}
\bibinfo{author}{{Hastings}, M.~B.} \& \bibinfo{author}{{Freedman}, M.~H.}
\newblock \bibinfo{title}{{Obstructions To Classically Simulating The Quantum
  Adiabatic Algorithm}}.
\newblock \emph{\bibinfo{journal}{arXiv e-prints}}
  \bibinfo{pages}{arXiv:1302.5733} (\bibinfo{year}{2013}).
\newblock \eprint{1302.5733}.

\bibitem{Richerme2013}
\bibinfo{author}{Richerme, P.} \emph{et~al.}
\newblock \bibinfo{title}{Experimental performance of a quantum simulator:
  Optimizing adiabatic evolution and identifying many-body ground states}.
\newblock \emph{\bibinfo{journal}{Phys. Rev. A}} \textbf{\bibinfo{volume}{88}},
  \bibinfo{pages}{012334} (\bibinfo{year}{2013}).

\bibitem{mcclean2016}
\bibinfo{author}{McClean, J.~R.}, \bibinfo{author}{Romero, J.},
  \bibinfo{author}{Babbush, R.} \& \bibinfo{author}{Aspuru-Guzik, A.}
\newblock \bibinfo{title}{The theory of variational hybrid quantum-classical
  algorithms}.
\newblock \emph{\bibinfo{journal}{New Journal of Physics}}
  \textbf{\bibinfo{volume}{18}}, \bibinfo{pages}{023023}
  (\bibinfo{year}{2016}).

\bibitem{Kokail2018}
\bibinfo{author}{{Kokail}, C.} \emph{et~al.}
\newblock \bibinfo{title}{{Self-Verifying Variational Quantum Simulation of the
  Lattice Schwinger Model}}.
\newblock \emph{\bibinfo{journal}{arXiv e-prints}}
  \bibinfo{pages}{arXiv:1810.03421} (\bibinfo{year}{2018}).
\newblock \eprint{1810.03421}.

\bibitem{harrow2016qaoa}
\bibinfo{author}{Farhi, E.} \& \bibinfo{author}{Harrow, A.~W.}
\newblock \bibinfo{title}{Quantum supremacy through the quantum approximate
  optimization algorithm}.
\newblock \emph{\bibinfo{journal}{arXiv preprint arXiv:1602.07674}}
  (\bibinfo{year}{2016}).

\bibitem{Lloyd2018}
\bibinfo{author}{{Lloyd}, S.}
\newblock \bibinfo{title}{{Quantum approximate optimization is computationally
  universal}}.
\newblock \emph{\bibinfo{journal}{arXiv e-prints}}
  \bibinfo{pages}{arXiv:1812.11075} (\bibinfo{year}{2018}).
\newblock \eprint{1812.11075}.

\bibitem{Preskill2018}
\bibinfo{author}{Preskill, J.}
\newblock \bibinfo{title}{Quantum {C}omputing in the {NISQ} era and beyond}.
\newblock \emph{\bibinfo{journal}{{Quantum}}} \textbf{\bibinfo{volume}{2}},
  \bibinfo{pages}{79} (\bibinfo{year}{2018}).

\bibitem{Zhou2018}
\bibinfo{author}{{Zhou}, L.}, \bibinfo{author}{{Wang}, S.-T.},
  \bibinfo{author}{{Choi}, S.}, \bibinfo{author}{{Pichler}, H.} \&
  \bibinfo{author}{{Lukin}, M.~D.}
\newblock \bibinfo{title}{{Quantum Approximate Optimization Algorithm:
  Performance, Mechanism, and Implementation on Near-Term Devices}}.
\newblock \emph{\bibinfo{journal}{arXiv e-prints}}
  \bibinfo{pages}{arXiv:1812.01041} (\bibinfo{year}{2018}).
\newblock \eprint{1812.01041}.

\bibitem{Hastings2019}
\bibinfo{author}{{Hastings}, M.~B.}
\newblock \bibinfo{title}{{Classical and Quantum Bounded Depth Approximation
  Algorithms}}.
\newblock \emph{\bibinfo{journal}{arXiv e-prints}}
  \bibinfo{pages}{arXiv:1905.07047} (\bibinfo{year}{2019}).
\newblock \eprint{1905.07047}.

\bibitem{bapat2018}
\bibinfo{author}{Bapat, A.} \& \bibinfo{author}{Jordan, S.}
\newblock \bibinfo{title}{Bang-bang control as a design principle for classical
  and quantum optimization algorithms}.
\newblock \emph{\bibinfo{journal}{arXiv preprint arXiv:1812.02746}}
  (\bibinfo{year}{2018}).

\bibitem{verdon2018}
\bibinfo{author}{Verdon, G.}, \bibinfo{author}{Pye, J.} \&
  \bibinfo{author}{Broughton, M.}
\newblock \bibinfo{title}{A universal training algorithm for quantum deep
  learning}.
\newblock \emph{\bibinfo{journal}{arXiv preprint arXiv:1806.09729}}
  (\bibinfo{year}{2018}).

\bibitem{Jiang2017}
\bibinfo{author}{Jiang, Z.}, \bibinfo{author}{Rieffel, E.~G.} \&
  \bibinfo{author}{Wang, Z.}
\newblock \bibinfo{title}{Near-optimal quantum circuit for grover’s
  unstructured search using a transverse field}.
\newblock \emph{\bibinfo{journal}{Physical Review A}}
  \textbf{\bibinfo{volume}{95}} (\bibinfo{year}{2017}).
\newblock \urlprefix\url{http://dx.doi.org/10.1103/PhysRevA.95.062317}.

\bibitem{Wang2018}
\bibinfo{author}{Wang, Z.}, \bibinfo{author}{Hadfield, S.},
  \bibinfo{author}{Jiang, Z.} \& \bibinfo{author}{Rieffel, E.~G.}
\newblock \bibinfo{title}{Quantum approximate optimization algorithm for
  maxcut: A fermionic view}.
\newblock \emph{\bibinfo{journal}{Physical Review A}}
  \textbf{\bibinfo{volume}{97}} (\bibinfo{year}{2018}).
\newblock \urlprefix\url{http://dx.doi.org/10.1103/PhysRevA.97.022304}.

\bibitem{brady2019}
\bibinfo{author}{Brady, L.}, \bibinfo{author}{Bapat, A.} \&
  \bibinfo{author}{Gorshkov, A.}
\newblock \emph{\bibinfo{journal}{in preparation}}  (\bibinfo{year}{2019}).

\bibitem{crooks2018}
\bibinfo{author}{Crooks, G.~E.}
\newblock \bibinfo{title}{Performance of the quantum approximate optimization
  algorithm on the maximum cut problem}.
\newblock \emph{\bibinfo{journal}{arXiv preprint arXiv:1811.08419}}
  (\bibinfo{year}{2018}).

\bibitem{Mbeng2019}
\bibinfo{author}{Mbeng, G.~B.}, \bibinfo{author}{Fazio, R.} \&
  \bibinfo{author}{Santoro, G.}
\newblock \bibinfo{title}{Quantum annealing: a journey through digitalization,
  control, and hybrid quantum variational schemes} (\bibinfo{year}{2019}).
\newblock \eprint{1906.08948}.

\bibitem{Ho2018a}
\bibinfo{author}{Ho, W.~W.} \& \bibinfo{author}{Hsieh, T.~H.}
\newblock \bibinfo{title}{{Efficient variational simulation of non-trivial
  quantum states}}.
\newblock \emph{\bibinfo{journal}{SciPost Phys.}} \textbf{\bibinfo{volume}{6}},
  \bibinfo{pages}{29} (\bibinfo{year}{2019}).

\bibitem{Ho2018b}
\bibinfo{author}{{Ho}, W.~W.}, \bibinfo{author}{{Jonay}, C.} \&
  \bibinfo{author}{{Hsieh}, T.~H.}
\newblock \bibinfo{title}{{Ultrafast State Preparation via the Quantum
  Approximate Optimization Algorithm with Long Range Interactions}}.
\newblock \emph{\bibinfo{journal}{arXiv e-prints}}
  \bibinfo{pages}{arXiv:1810.04817} (\bibinfo{year}{2018}).

\bibitem{Koffel2012}
\bibinfo{author}{Koffel, T.}, \bibinfo{author}{Lewenstein, M.} \&
  \bibinfo{author}{Tagliacozzo, L.}
\newblock \bibinfo{title}{Entanglement entropy for the long-range ising chain
  in a transverse field}.
\newblock \emph{\bibinfo{journal}{Physical Review Letters}}
  \textbf{\bibinfo{volume}{109}} (\bibinfo{year}{2012}).

\bibitem{Kim2009}
\bibinfo{author}{Kim, K.} \emph{et~al.}
\newblock \bibinfo{title}{Entanglement and tunable spin-spin couplings between
  trapped ions using multiple transverse modes}.
\newblock \emph{\bibinfo{journal}{Physical Review Letters}}
  \textbf{\bibinfo{volume}{103}} (\bibinfo{year}{2009}).

\bibitem{Pagano2019}
\bibinfo{author}{Pagano, G.} \emph{et~al.}
\newblock \bibinfo{title}{Cryogenic trapped-ion system for large scale quantum
  simulation}.
\newblock \emph{\bibinfo{journal}{Quantum Science and Technology}}
  \textbf{\bibinfo{volume}{4}}, \bibinfo{pages}{014004} (\bibinfo{year}{2019}).

\bibitem{Porras2004}
\bibinfo{author}{Porras, D.} \& \bibinfo{author}{Cirac, J.~I.}
\newblock \bibinfo{title}{{Effective Quantum Spin Systems with Trapped Ions}}.
\newblock \emph{\bibinfo{journal}{Phys. Rev. Lett.}}
  \textbf{\bibinfo{volume}{92}}, \bibinfo{pages}{207901}
  (\bibinfo{year}{2004}).

\bibitem{Jaschke2018}
\bibinfo{author}{Jaschke, D.}, \bibinfo{author}{Wall, M.~L.} \&
  \bibinfo{author}{Carr, L.~D.}
\newblock \bibinfo{title}{Open source matrix product states: Opening ways to
  simulate entangled many-body quantum systems in one dimension}.
\newblock \emph{\bibinfo{journal}{Computer Physics Communications}}
  \textbf{\bibinfo{volume}{225}}, \bibinfo{pages}{59 -- 91}
  (\bibinfo{year}{2018}).

\bibitem{dylewsky2016}
\bibinfo{author}{Dylewsky, D.}, \bibinfo{author}{Freericks, J.},
  \bibinfo{author}{Wall, M.}, \bibinfo{author}{Rey, A.} \&
  \bibinfo{author}{Foss-Feig, M.}
\newblock \bibinfo{title}{Nonperturbative calculation of phonon effects on spin
  squeezing}.
\newblock \emph{\bibinfo{journal}{Physical Review A}}
  \textbf{\bibinfo{volume}{93}}, \bibinfo{pages}{013415}
  (\bibinfo{year}{2016}).

\bibitem{Hadfield2018}
\bibinfo{author}{Hadfield, S.}
\newblock \bibinfo{title}{Quantum algorithms for scientific computing and
  approximate optimization}.
\newblock \emph{\bibinfo{journal}{arXiv preprint arXiv:1805.03265}}
  (\bibinfo{year}{2018}).

\bibitem{Bremner2016}
\bibinfo{author}{Bremner, M.~J.}, \bibinfo{author}{Montanaro, A.} \&
  \bibinfo{author}{Shepherd, D.~J.}
\newblock \bibinfo{title}{Average-case complexity versus approximate simulation
  of commuting quantum computations}.
\newblock \emph{\bibinfo{journal}{Phys. Rev. Lett.}}
  \textbf{\bibinfo{volume}{117}}, \bibinfo{pages}{080501}
  (\bibinfo{year}{2016}).

\bibitem{Bouland2019}
\bibinfo{author}{Bouland, A.}, \bibinfo{author}{Fefferman, B.},
  \bibinfo{author}{Nirkhe, C.} \& \bibinfo{author}{Vazirani, U.}
\newblock \bibinfo{title}{On the complexity and verification of quantum random
  circuit sampling}.
\newblock \emph{\bibinfo{journal}{Nat. Phys.}} \textbf{\bibinfo{volume}{15}},
  \bibinfo{pages}{159--163} (\bibinfo{year}{2019}).

\bibitem{Wang2015}
\bibinfo{author}{{Wang}, S.-T.} \& \bibinfo{author}{{Duan}, L.-M.}
\newblock \bibinfo{title}{{Certification of Boson Sampling Devices with
  Coarse-Grained Measurements}}.
\newblock \emph{\bibinfo{journal}{arXiv e-prints}}
  \bibinfo{pages}{arXiv:1601.02627} (\bibinfo{year}{2016}).
\newblock \eprint{1601.02627}.

\bibitem{Kandala2017}
\bibinfo{author}{Kandala, A.} \emph{et~al.}
\newblock \bibinfo{title}{Hardware-efficient variational quantum eigensolver
  for small molecules and quantum magnets}.
\newblock \emph{\bibinfo{journal}{Nature}} \textbf{\bibinfo{volume}{549}},
  \bibinfo{pages}{242 EP --} (\bibinfo{year}{2017}).

\bibitem{Hempel2018}
\bibinfo{author}{Hempel, C.} \emph{et~al.}
\newblock \bibinfo{title}{Quantum chemistry calculations on a trapped-ion
  quantum simulator}.
\newblock \emph{\bibinfo{journal}{Phys. Rev. X}} \textbf{\bibinfo{volume}{8}},
  \bibinfo{pages}{031022} (\bibinfo{year}{2018}).

\bibitem{Nam2019}
\bibinfo{author}{{Nam}, Y.} \emph{et~al.}
\newblock \bibinfo{title}{{Ground-state energy estimation of the water molecule
  on a trapped ion quantum computer}}.
\newblock \emph{\bibinfo{journal}{arXiv e-prints}}
  \bibinfo{pages}{arXiv:1902.10171} (\bibinfo{year}{2019}).
\newblock \eprint{1902.10171}.

\bibitem{Rigetti2017}
\bibinfo{author}{{Otterbach}, J.~S.} \emph{et~al.}
\newblock \bibinfo{title}{{Unsupervised Machine Learning on a Hybrid Quantum
  Computer}}.
\newblock \emph{\bibinfo{journal}{arXiv e-prints}}
  \bibinfo{pages}{arXiv:1712.05771} (\bibinfo{year}{2017}).
\newblock \eprint{1712.05771}.

\bibitem{Giovannetti2011}
\bibinfo{author}{Giovannetti, V.}, \bibinfo{author}{Lloyd, S.} \&
  \bibinfo{author}{Maccone, L.}
\newblock \bibinfo{title}{Advances in quantum metrology}.
\newblock \emph{\bibinfo{journal}{Nature Photonics}}
  \textbf{\bibinfo{volume}{5}}, \bibinfo{pages}{222 EP --}
  (\bibinfo{year}{2011}).

\bibitem{Farhi2014maxE3lin2}
\bibinfo{author}{{Farhi}, E.}, \bibinfo{author}{{Goldstone}, J.} \&
  \bibinfo{author}{{Gutmann}, S.}
\newblock \bibinfo{title}{{A Quantum Approximate Optimization Algorithm Applied
  to a Bounded Occurrence Constraint Problem}}.
\newblock \emph{\bibinfo{journal}{arXiv e-prints}}
  \bibinfo{pages}{arXiv:1412.6062} (\bibinfo{year}{2014}).
\newblock \eprint{1412.6062}.

\bibitem{Wineland1998}
\bibinfo{author}{Wineland, D.} \emph{et~al.}
\newblock \bibinfo{title}{Experimental issues in coherent quantum-state
  manipulation of trapped atomic ions}.
\newblock \emph{\bibinfo{journal}{J. Res. Natl. Inst. Stand. Technol.}}
  \textbf{\bibinfo{volume}{103}}, \bibinfo{pages}{259--328}
  (\bibinfo{year}{1998}).

\bibitem{Olmschenk2007}
\bibinfo{author}{Olmschenk, S.} \emph{et~al.}
\newblock \bibinfo{title}{Manipulation and detection of a trapped
  ${\mathrm{yb}}^{+}$ hyperfine qubit}.
\newblock \emph{\bibinfo{journal}{Phys. Rev. A}} \textbf{\bibinfo{volume}{76}},
  \bibinfo{pages}{052314} (\bibinfo{year}{2007}).

\bibitem{Brown2004}
\bibinfo{author}{Brown, K.~R.}, \bibinfo{author}{Harrow, A.~W.} \&
  \bibinfo{author}{Chuang, I.~L.}
\newblock \bibinfo{title}{Arbitrarily accurate composite pulse sequences}.
\newblock \emph{\bibinfo{journal}{Phys. Rev. A}} \textbf{\bibinfo{volume}{70}},
  \bibinfo{pages}{052318} (\bibinfo{year}{2004}).

\bibitem{Molmer1999}
\bibinfo{author}{S\o{}rensen, A.} \& \bibinfo{author}{M\o{}lmer, K.}
\newblock \bibinfo{title}{Quantum computation with ions in thermal motion}.
\newblock \emph{\bibinfo{journal}{Phys. Rev. Lett.}}
  \textbf{\bibinfo{volume}{82}}, \bibinfo{pages}{1971--1974}
  (\bibinfo{year}{1999}).

\bibitem{James1998}
\bibinfo{author}{James, D. F.~V.}
\newblock \bibinfo{title}{Quantum dynamics of cold trapped ions with
  application to quantum computation}.
\newblock \emph{\bibinfo{journal}{Applied Physics B}}
  \textbf{\bibinfo{volume}{66}}, \bibinfo{pages}{181} (\bibinfo{year}{1998}).

\end{thebibliography}

\section{Supplementary material}

\subsection{Quantum Approximate Optimization Algorithm (QAOA)}
\label{sec:qaoa}

The QAOA is an approximate optimization algorithm first introduced in 2014 by Farhi \emph{et al.} \cite{Farhi2014}, and has since enjoyed growing interest. The QAOA uses alternating evolutions under two non-commuting operators, typically a problem (or cost) Hamiltonian $H_A$ that encodes the cost function on the diagonal in (say) the $\sigma^{x}$ basis, and a transverse term $H_B = -\suml{i=0}{N}{\sigma^{y}_i}$ that generates transitions between bit strings, such that the initial state $\ket{+}_y^{\otimes N}$ evolves into an approximate ground state of $H_A$. 

Practically, the most valuable feature of the QAOA seems to be its ``learnability'' via a classical outer loop optimizer, where the discovery of the evolution angles in the optimal QAOA schedule is achieved via the discovery of structure in the angle sequences \cite{Zhou2018, brady2019, crooks2018}. These patterns are seen quite generally across local Hamiltonian problems, and while steps towards a theory describing optimal QAOA sequences have been taken \cite{brady2019}, several questions surrounding it remain open. Regardless, the structure in optimal QAOA schedules may be harnessed to implement approximate state preparation in a scalable manner and with a low overhead on quantum resources. We present a new heuristic method that helps achieves this goal.  


First, we discuss how to discover optimal QAOA1 schedules, i.e., QAOA schedules for $p=1$.

\subsubsection{QAOA, $p=1$} \label{sec:QAOA1}
Despite its apparent simplicity, the $p=1$ QAOA (or QAOA1) can be a powerful state preparation ansatz. For example, hardness-of-sampling results are known for QAOA1 circuits \cite{harrow2016qaoa}, closely mirroring the hardness of sampling from instantaneous quantum polynomial (IQP) circuits (see next section for details). Furthermore, it is known that the performance of the QAOA1 for certain combinatorial optimization problems can be competitive with the best classical algorithms for the same problems \cite{Farhi2014maxE3lin2}. 
Another desirable feature of the QAOA1 for local spin Hamiltonians is the tractability of computing energy expectation values, as observed in \cite{Farhi2014}. A very similar result has also been known in the setting of quantum dynamics \cite{dylewsky2016, Hadfield2018}. For a two-local transverse field spin Hamiltonian as in Eq. (1) in the main text, this leads to a formula for the energy expectation under a state produced by the QAOA1, starting from the product state $\ket{+}^{\otimes N}$. These formulas are applicable to many cases of interest in quantum state preparation and optimization. Importantly, the time complexity to compute the formula is $O(N^3)$ in the worst case, making it tractable to optimize the QAOA1 protocols for large spin chains. 

\renewcommand{\figurename}{\textbf{Figure}}
\renewcommand{\thefigure}{S1}
\begin{figure*}[t]
\centering
\includegraphics[clip=true, trim=10pt 0 10pt 0,width=0.49\textwidth]{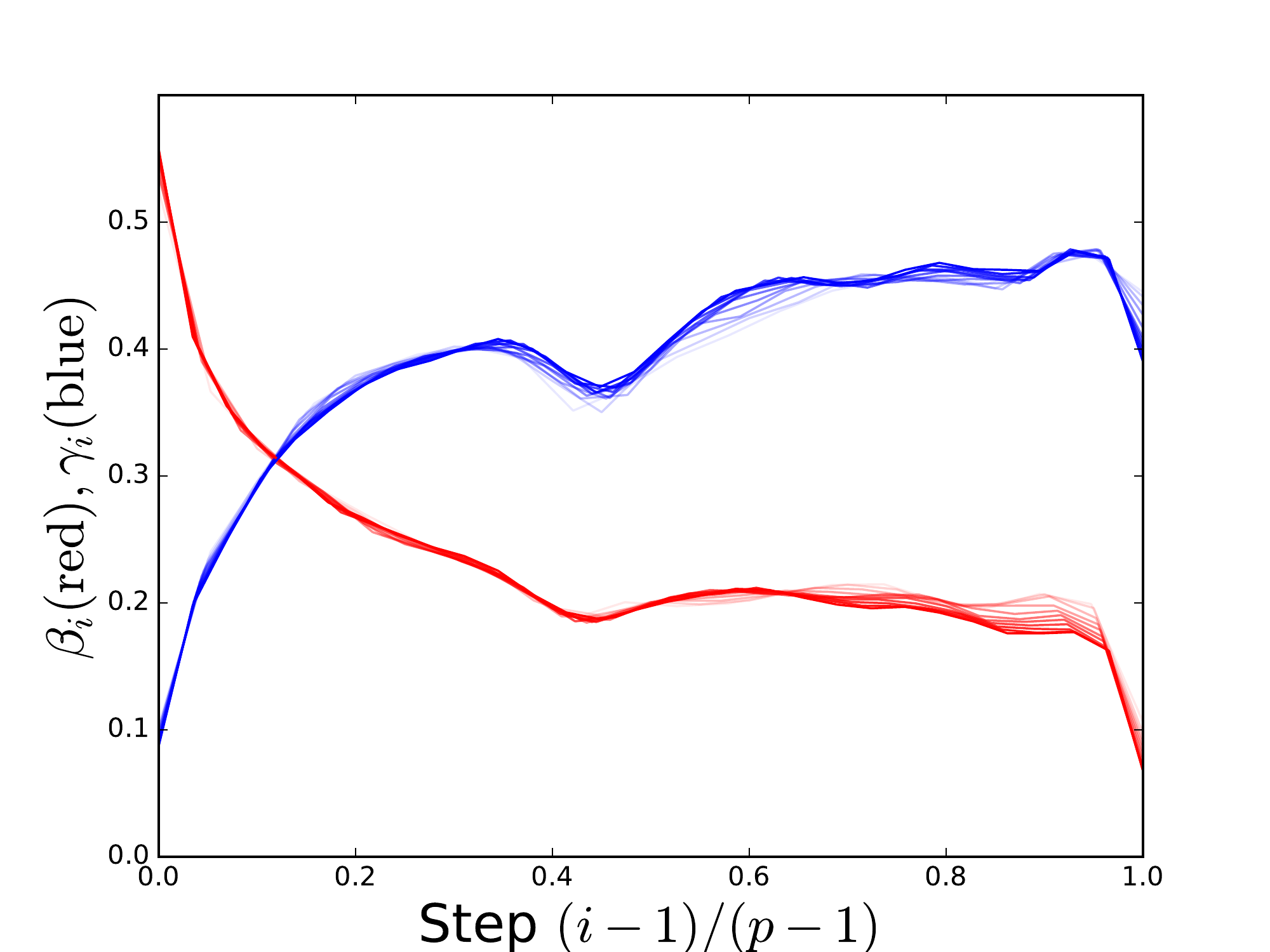}\includegraphics[clip=true, trim=10pt 0 10pt 0,width=0.49\textwidth]{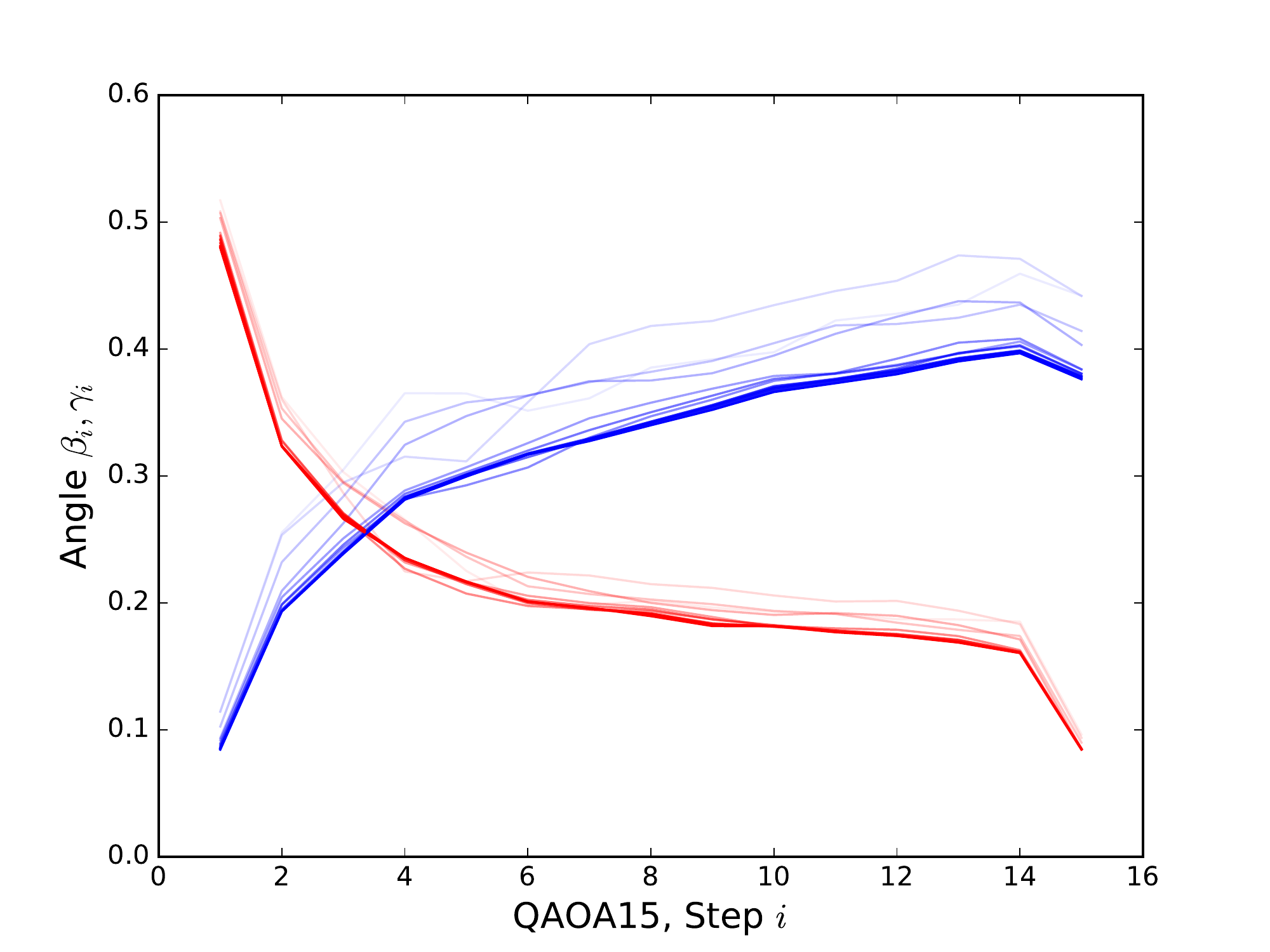} \caption{{\bf Convergence in $p$ and $N$.} Convergence of optimal angle curves with increasing QAOA layers $p$ (left), and number of spins $N$ (right). The $p$-convergence plot was generated for an $N=8$ spin system, for $p$ ranging from 20 up to 30, with higher $p$ shaded darker. The $N$-convergence figure was generated for a $15$ layer QAOA, for $N$ in the range of $4$ to $14$, with higher $N$ curves shaded darker.}
\label{fig:CurveScalingVpAndN}
\end{figure*}

\subsubsection{QAOA, $p>1$}\label{sec:QAOAp}

The general analytical formula for $p=1$ does not extend to the case where we apply the QAOA for more than one layer. Here, we must turn to classical numerical methods to find the optimal QAOA angles $\beta_i, \gamma_i$ for each layer $i$. For $p$ layers, this is an optimization on a $2p$-dimensional space that grows exponentially with the depth of the circuit. However, numerics done here and in \cite{crooks2018,Zhou2018} have identified the existence of minima that exhibit patterns in the optimal QAOA angles, namely that the angles, when plotted as a function of their index $i$, form smooth curves for any $p$. While this observation points to a deeper theoretical mechanism at play, it does not directly simplify the optimization problem, since we must still search over all approximately smooth sequences of the angles. Zhou \emph{et al.} \cite{Zhou2018} have exploited the smoothness of the functions by carrying out searches in the Fourier domain. Here, we follow a different route that arises from some novel observations of these family of minima.

For each $p$, denote the special optimal angles by $\curly{\paren{\boldsymbol{\beta}^{*(p)},\boldsymbol{\gamma}^{*(p)}}}_p$, which we can also think of as a pair of angle curves (as a function of step index $i$). As $p$ is varied, we may think of these minima as a family. We numerically find that this family exhibits the following desirable features (for $p$ sufficiently large): 
\begin{enumerate}
    \item The angles are non-negative, small and bounded. 
    \item For $p$ sufficiently large, the two angle sequences ${\beta}^{*(p)}$ and ${\gamma}^{*(p)}$ are approximately smooth.
    \item The angle sequence ${\beta}^{*(p)}$ (and correspondingly, ${\gamma}^{*(p)}$) when viewed as a function on the normalized time parameter $s_i = \frac{i-1}{p-1}$, is convergent in the parameter $p$. In other words, as $p$ is increased, the angle sequences ${\beta}^{*(p)}$ and ${\gamma}^{*(p)}$ approach a smooth, asymptotic curve (See Fig.~\ref{fig:CurveScalingVpAndN}.) 
    \item The energy expectation $E(\beta^{*(p)},\gamma^{*(p)})$ approaches the global minimum as $p \rightarrow \infty$, and hence this family is asymptotically optimal.
\end{enumerate}
The significance of the first point is that in experimental settings, large evolution times are infeasible to implement due to decoherence, so these minima correspond to practicable QAOA protocols. The third and fourth points suggest an inductive algorithm where a locally optimal schedule for a given $p$ may be discovered using the optimal schedule for $p-1$ as a prior.

Point 3 in the above list is a novel observation that allows us to construct a heuristic that is efficiently scalable for large $p$. The main idea behind this construction is that the minimal angle curves for a larger $p$ may be guessed from the optimal curve of a smaller $p' < p$ by interpolation.

Using the above points, we use a bootstrapping algorithm to find the optimal angle sequences, ${\beta}^{*(p)}$ and ${\gamma}^{*(p)}$, for a given $p$, as described below. Let $q=1,\ldots, p$ denote an intermediate angle index. Then:
\begin{enumerate}
\item For $q=1$, use an analytic formula to find ${\beta}^{*(1)}$ and ${\gamma}^{*(1)}$.
\item For $q=2$, choose an initial guess of ${\beta}^{(2)} = \left({\beta}^{*(1)},{\beta}^{*(1)}-0.2\right)$ and ${\gamma}^{(2)} = \left({\gamma}^{*(1)},{\gamma}^{*(1)}+0.2\right)$.
\item Perform a local optimization of $\beta^{(2)}$ and $\gamma^{(2)}$ in order to find $\beta^{*(2)}$ and $\gamma^{*(2)}$.
\item Repeat the next steps (5-7) for $q = 3,\ldots, p$.
\item Create interpolating functions through the angle sequences, $\beta^{*(q-1)}$ and $\gamma^{*(q-1)}$, using the normalized time $s_i = \frac{i-1}{q-2}$ as the independent parameter (we use a linear interpolation for $q=3$ and cubic for $q>3$).
\item Choose the initial guesses for $\beta^{(q)}$ and $\gamma^{(q)}$ by sampling the interpolating function from (5) at evenly spaced points separated by a normalized time distance of $\Delta s = 1/(q-1)$.
\item Perform a local optimization of $\beta^{(q)}$ and $\gamma^{(q)}$ in order to find $\beta^{*(q)}$ and $\gamma^{*(q)}$.
\end{enumerate}

The resulting angles $\beta^{*(p)}$ and $\gamma^{*(p)}$ should be at least a good local minimum of the energy expectation value and approaches the global minimum as $p\to\infty$.

The $q=2$ interpolation in step 2 is based on our observation that the $\beta$ angles tend to curve down at the end and the $\gamma$ angles tend to curve up.

An important feature of our algorithm is that its asymptotic runtime is expected to be efficient in $p$. This feature is predicated on the previous result that the angle curves are generally convergent as $p$ tends to infinity. The argument proceeds as follows: if we assume a maximal deviation of the initial guess for layer $q$ to be $\epsilon_q \ge 0$, then the total $l_2$-norm distance between the initial guess and the optimized curve is no greater than $\epsilon_q\sqrt{q}$, by the Cauchy-Schwarz inequality. Therefore, the local search algorithm is confined to a ball of radius at most $\epsilon_q\sqrt{q}$, and for a fixed error tolerance, the convergence time for a standard local optimizer is $O(\epsilon_q^2q)$. Summing over convergence times for all from $q=1,\ldots, p$, we have
\begin{equation}
  \label{eq:convergence}
  T = O\left(\suml{q=1}{p}{q\epsilon_q^2}\right) \le O(p^2)
\end{equation}

The last inequality above comes about as follows: while the summand depends on the convergence rate of the sequence $\curly{\epsilon_q}_{q=1}^p$, it is upper bounded by $O(q)$ for a converging set of paths and an initial error $\epsilon_1$ of order 1. The latter is true since our angle search domain is bounded and independent of $N$. Therefore, the sum is no greater than $O(p^2)$. In practice, even faster runtimes are possible. Therefore, the bootstrap algorithm exploits the structure of the special minima and provides a scalable route to multi-step QAOA for the long-range TFIM. In fact, as discussed in the supplement and in \cite{brady2019}, there is mounting numerical evidence that the path approach applies across a very general variety of models on discrete as well as continuous systems.
\renewcommand{\figurename}{\textbf{Figure}}
\renewcommand{\thefigure}{S2}
\begin{figure*}[t]
\centering
\includegraphics[clip=true, trim=5pt 45pt 10pt 40pt,width=0.95\textwidth]{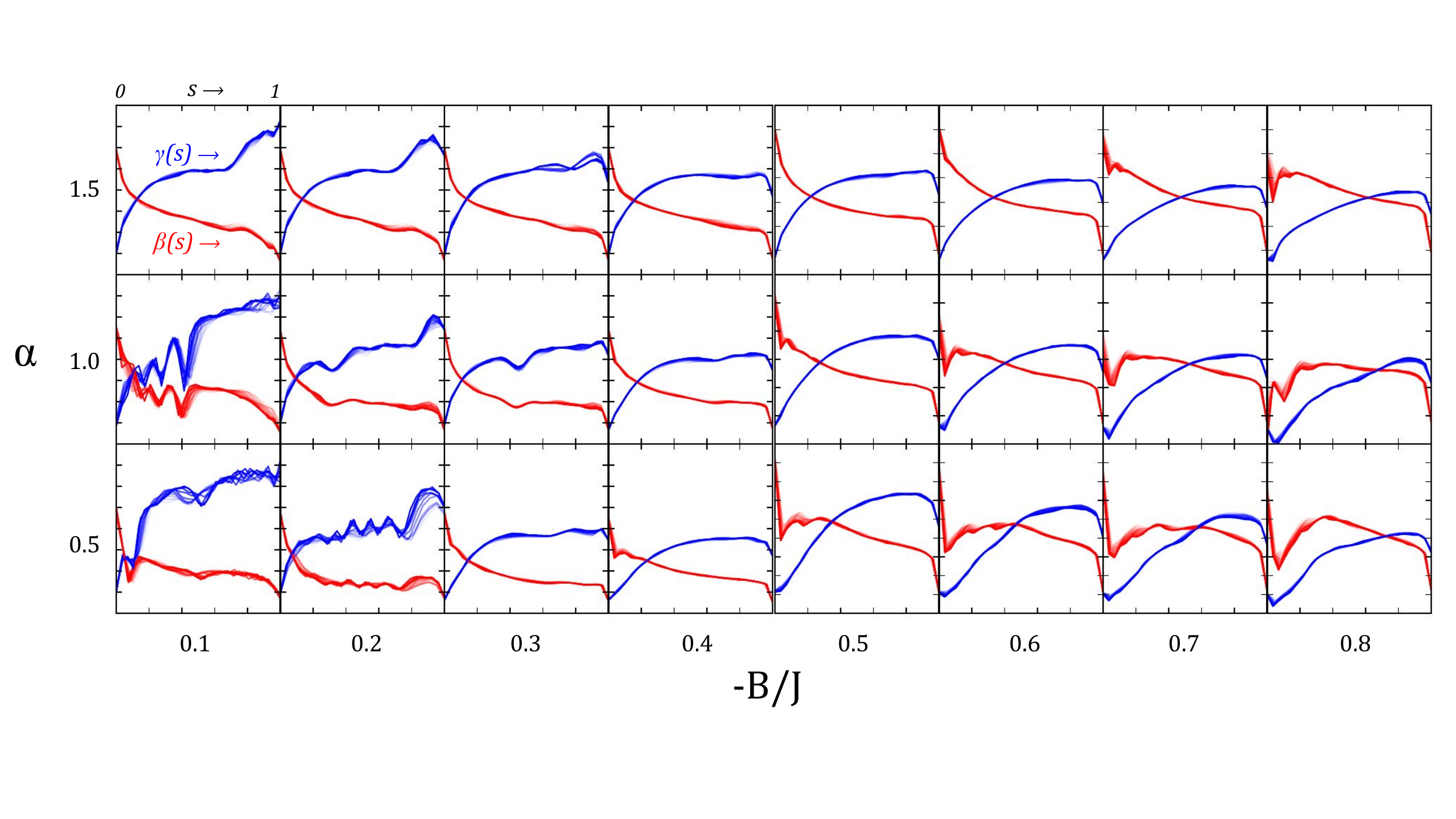}
\caption{{\bf Angle sequence curves.} A collage of angle sequence curves, arranged by the Hamiltonian parameters for which they were computed. In each subplot, curves for different $p$ ranging from 20 to 30 are overlaid, with higher $p$ curves shaded darker. The horizontal axis represents fractional step $s = (i-1)/(p-1)$ ranging from 0 to 1, while the vertical axis gives the value of the angles $\beta$ (red), and $\gamma$ (blue) in the range $[0,0.6]$. The subplots are arranged horizontally by $-B/J_0$, increasing from 0.1 to 0.8 in steps of 0.1 (from left to right), and vertically by the long-range power $\alpha=0.5,1.0,1.5$ (bottom to top). This collage shows the persistence of structure in the optimal angle sequences for a range of Hamiltonians within the same family.}
\label{fig:CurveGrid}
\end{figure*}

\subsubsection{Convergence in $N$}\label{sec:largen}

In the previous sections, we introduced a bootstrap algorithm that is asymptotically efficient in the number of layers $p$. However, in order to be fully scalable the algorithm must also be scalable in the system size $N$. This may not be possible in general (say for random spin models), as the optimized angles for a particular small system may have no bearing on the angles for a larger system. However, for the long-range TFIM, and indeed any translationally-invariant model with a well-defined notion of metric and dimension arising from the functional form of the coupling coefficients $J_{ij}$, it is reasonable to expect that the optimized angles depend on system size in a predictable way. This is indeed the case for the long-range TFIM. There, it can be seen that the angle curves for varying $N$ appear similar in shape. Usefully, the curves also appear to be \emph{convergent} to an idealized curve for a hypothetical continuous, long-range spin chain. Once again, this feature suggests that the optimized QAOA angle curves for small systems may be used as initial guesses for larger systems \emph{within the same Hamiltonian family}.

While it is not clear (due to numerical limitations) how fast the curves converge, we argue that the rate should be weakly dependent (or independent) of the system size $N$. For a given coupling function (such as inverse power-law) that decays as a function of distance, we define a characteristic length scale, which may be called the \emph{skin depth} $\delta$, that is the number of sites from the boundary that the coupling is a factor of $e$ smaller than the nearest-neighbour value. In other words, we define $\delta$ such that $J_{i,i+\delta} \sim J_{i,i+1}/e$. Clearly, $\delta$ is independent of the system size $N$ and depends only on the parameters of the coupling function. For instance, for the long-range TFIM, $\delta \sim e^{1/\alpha}$. As $N$ tends to infinity, the \emph{fractional} skin depth $\delta/N$ then ``falls away'' and becomes vanishing with respect to the bulk region of the chain. Now, we make the assumption that any deviations in the optimal QAOA schedules from $N$ to $N+1$ arise from change in the fractional skin depth, which is reasonable for a translationally invariant model. The incremental change in the fractional skin depth from $N$ to $N+1$ is $\delta/N-\delta/(N+1) \sim O(1/N^2)$. Therefore, if the change in the optimal QAOA curves $\epsilon_N$ (in, say, $l_1$-norm distance) is a smooth function of the the fractional skin depth, then we expect it to vary as $\epsilon_N \sim 1/\text{poly}(N)$. Therefore, the total running time of a bootstrap from small system sizes to a given size $N$ should be $O\paren{\suml{k=1}{N}{1/\text{poly}(N)}}$ which is sub-linear in $N$.     
Combining this observation with the convergence in $p$, we see that for a given Hamiltonian family, optimized QAOA angle curves for small $p$ may be used as a rubric for the optimization for longer circuit depths. Furthermore, if the Hamiltonian is translationally-invariant with decaying interactions, the optimized QAOA schedules are expected to scale with $N$ as well. Therefore, the state preparation procedure under the QAOA for such a Hamiltonian family is scalable in circuit ``volume'', for a wide range of Hamiltonian parameters (Fig.~\ref{fig:CurveGrid}). This is our main theoretical contribution in this work. 

\renewcommand{\figurename}{\textbf{Figure}}
\renewcommand{\thefigure}{S3}
\begin{figure*}[t]
\centering
\includegraphics[clip=true, trim=5pt 0pt 10pt 40pt,width=0.49\textwidth]{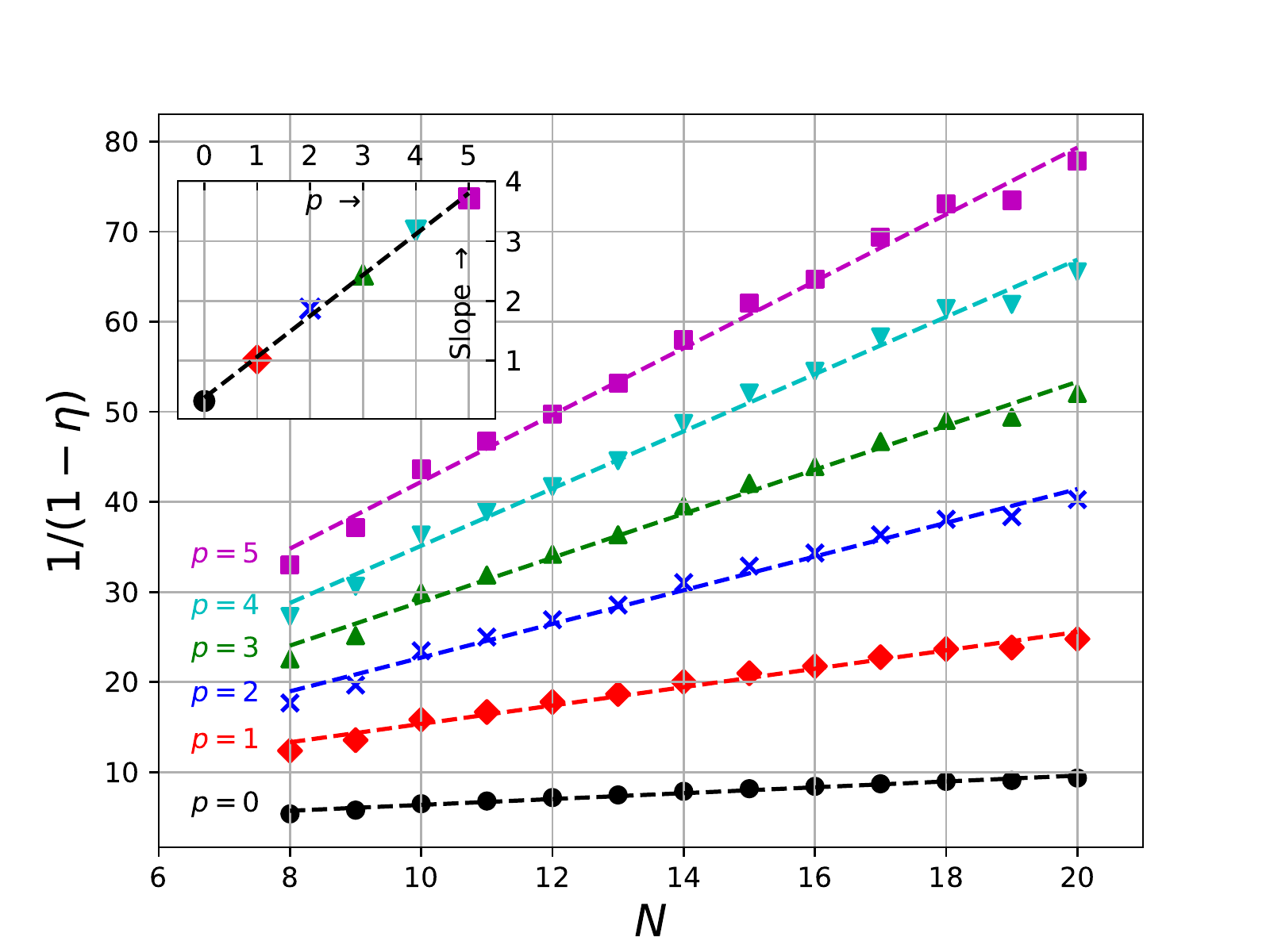}
\includegraphics[clip=true, trim=5pt 0pt 10pt 40pt,width=0.49\textwidth]{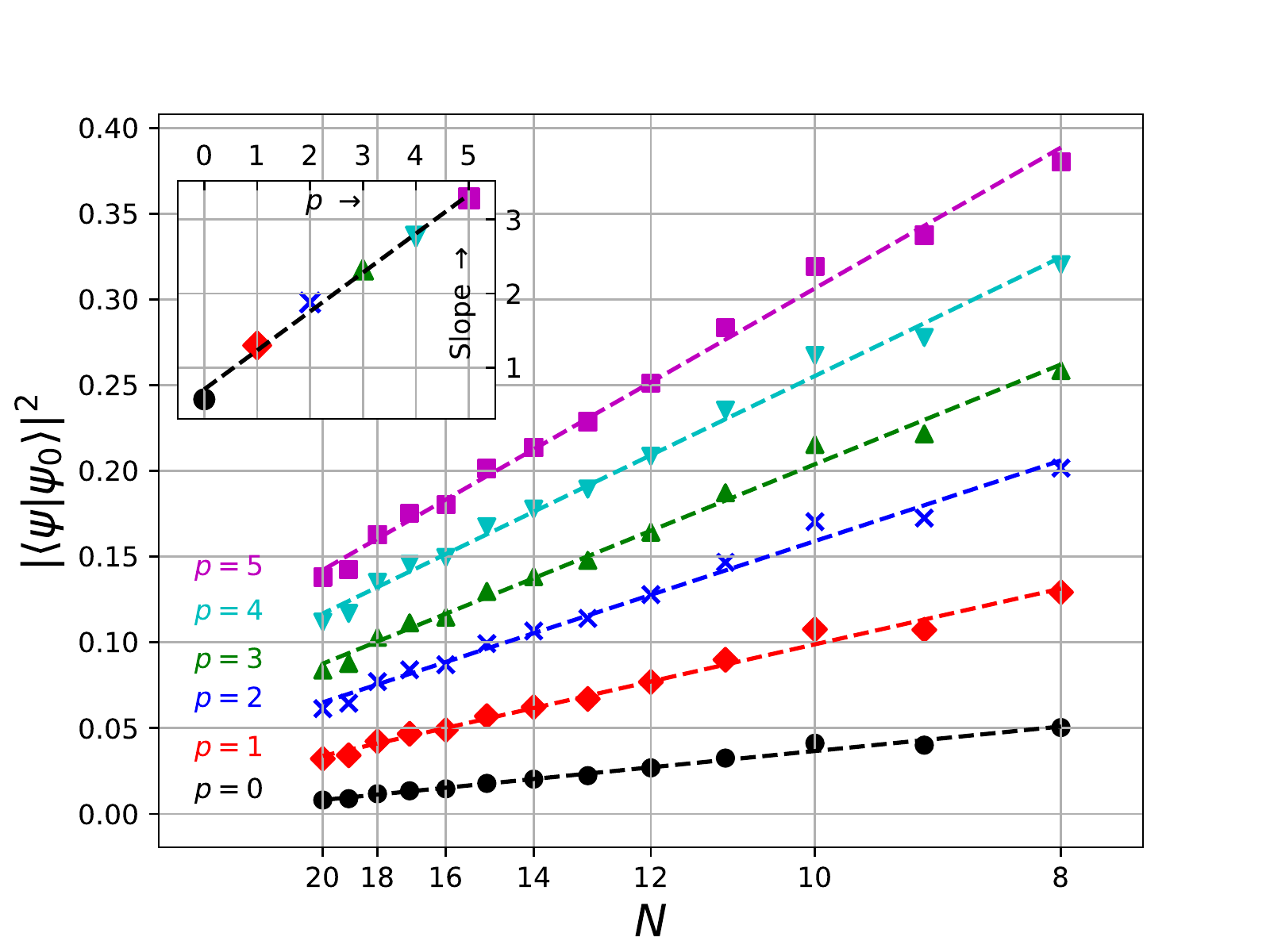}
\caption{{\bf Performance scaling in $p,N$.} Behaviour of performance parameters $\eta$ (left) and squared ground state overlap (right) with increasing number of spins $N$ ($x$ axis) and $p$ (colors), for ideal power-law coupling with $\alpha=1.1$. We find that for each $p$, $1/(1-\eta)$ grows linearly in $N$ with a slope that depends on $p$. (Inset) The slope is linear in $p$, suggesting that the performance converges to 1 as $\eta \sim 1-1/(pN)$. On the right, we empirically observe that $|\langle \psi|\psi_0\rangle|^2 \sim p/N$, indicating that constant overlap with the ground state can be achieved with linear depth QAOA. The $x$ axis has been scaled as $1/N$ so that the linear relationship with the squared overlap is apparent. The inset shows the linear trend with $p$.}
\label{fig:eta_vs_pN}
\end{figure*}

\red{
\subsubsection{Scaling of $\eta$ in $p, N$}\label{sec:largen}

Our performance parameter $\eta$, defined as 
\begin{equation}
\eta\equiv \frac{E(\vec{\beta},\vec{\gamma})-E_{max}}{E_{gs}-E_{max}},
\label{eq_eta}
\end{equation}
measures how close (in energy) the prepared state is to the ground state of the system. As described in previous sections, the optimal angle curves for QAOA appear to converge to a smooth, hypothetical curve, as a function of $p$ as well as $N$. We show that under the assumption that such a curve exists, there is a fast heuristic for finding optimal angles for any finite $p$ that is time-efficient in $p$ and the number of spins $N$ (when used in conjunction with the quantum device). In this section, we show that not only is the search efficient, but the quality of the optimum is numerically seen to improve with $p,N$ as well. 

In Fig.~\ref{fig:eta_vs_pN}, we show the result of the numerical study. We chose as the target Hamiltonian an idealized transverse field Ising model with inverse power-law couplings, with the power $\alpha=1.1$ chosen to closely mimic the experimental Hamiltonian. The number of spins was varied from $N=8$ to $20$. Via DMRG, the critical value of the transverse field for a finite chain can be located by maximizing the von Neumann entropy at half-cut. This was done independently for each value of $N$. Then, using our heuristic, we located the optimal angle curve, and computed $\eta$ for the final state prepared using this angle sequence, for each $N$. The plot shows the trend of $1/(1-\eta)$ with $N$, for a range of $p=0,1,2,3,4,5$, with $0$ corresponding to a trivial protocol where the initial state is returned. While the number of spins could not be extended beyond $20$ due to computational limitations, the trend is clear. We see that $1/(1-\eta)$ grows linearly with $N$ and $p$ (inset). While the linear trend in $N$ is encouraging, we similarly expect the inverse spectral gap (and indeed, the density of low-lying states) to increase with $N$. Empirically for the target Hamiltonian, we observe a gap scaling of $\sim 1/N^2$. Assuming the density of low-lying states scales similarly, this suggests that the squared overlap with the ground state should fall off with $N$. Numerics confirm this expectation and indicate a scaling of the squared overlap of $|\langle \psi| \psi_0\rangle|^2 \sim p/N$.

The linear scaling with $p$ for both the energy and fidelity metric, combined with a polynomial-time search heuristic, suggests that for any desired energy (or probability) threshold $\epsilon$, our approach allows us to approximate the state to within $1-\epsilon$ (in energy or fidelity) in time and number of layers that scale as $\text{poly}(N,1/\epsilon)$.

\subsubsection{Characteristic scale for $\eta$}\label{sec:largen}

The figure of merit $\eta$ characterizes how close the final state is to the ground state of the system. At $\eta=0$, the system is in the highest excited configuration, while $\eta=1$ corresponds to a perfectly prepared ground state. QAOA, starting from the initial state $\ket{+}^{\otimes n}$, gives a state with figure of merit $\eta \in [0,1]$, from the initial value of $\eta_0$. The difference between the final $\eta$ and $\eta_0$ indicate the success of our QAOA protocol. 

While $\eta$ is normalized to the range $[0,1]$, differences in $\eta$ are still somewhat arbitrary. In long-range Ising models with a transverse field $\eta_0$ is not $0$ but typically greater than $0.5$, making the difference in $\eta$ an unsatisfactory metric of success. Therefore, in addition to the initial and final $\eta$, we must provide a characteristic scale for $\eta$ that quantifies the typical deviation from $\eta_0$. A natural choice is the standard deviation of $\eta$ for QAOA with random angles. 

For QAOA1 with evolution angles $\beta, \gamma$, it is possible to estimate the standard deviation analytically  \blue{as a function of the underlying model parameters $B$ and $J_0$ and on the number of qubits $N$}. This derives from the analytical formula for the energy expectation $E(\beta, \gamma)$ which can be stated as follows:
\begin{equation}
E(\beta,\gamma) = E_{I} + E_{II} + E_{III}
\end{equation}
where
\begin{align}
&E_I = B\suml{i=1}{N}{\prodl{k\ne i}{}{\cos\paren{2\gamma J_{ik}}}}\\
&E_{II} = -\frac{\sin\paren{4\beta}}{2}\suml{i,j}{}{J_{ij}\sin\paren{2\gamma J_{ij}}\prodl{k\ne i,j}{}{\cos\paren{2\gamma J_{ik}}}}\\
&E_{III} =  - \frac{\sin^2\paren{2\beta}}{4}\suml{s=\pm1,i,j}{}{J_{ij}\prodl{k\ne i,j}{}{\cos\paren{2\gamma \paren{J_{ik}+(-1)^sJ_{jk}}}}}
\end{align}
where the Hamiltonian has long-range power law couplings $J_{ij}\sim \frac{1}{|i-j|^\alpha}$ (with $J_{ii}=0$), and a transverse field of strength $B$. Then, our goal is to compute the standard deviation (normalized by the spectral bandwidth $\Delta := E_{max} - E_{gs}$),
\begin{equation}
   \frac{\sigma_E}{\Delta} = \frac{\sqrt{\langle E^2\rangle_{\beta,\gamma} - \langle E\rangle_{\beta,\gamma}^2}}{\Delta}
\end{equation}
which gives us the characteristic scale for $\eta$. We define the average $\langle\cdot\rangle_{\beta,\gamma}$ as
\begin{equation}
\langle f \rangle_{\beta,\gamma} :=\lim_{T_{\beta},T_{\gamma}\rightarrow \infty} \frac{1}{4T_\beta T_\gamma}\intl{-T_\beta}{T_\beta}{\intl{-T_\gamma}{T_{\gamma}}{f(\beta,\gamma)d\beta d\gamma}}
\end{equation}
In the limit, the average is precisely the constant term of the Fourier transform of $f$. Since the function is a sum of trigonometric monomials, its moments over the angle variables $\beta, \gamma$ can be computed analytically term by term. We will need the following properties of the coupling function:
\begin{enumerate}
\item (Symmetry) Since the inverse power law only depends on distance between nodes, we have $J_{ij} = J_{(2j-i)j}$ In other words, the inverse power-law is symmetric under a lateral flip (or ``mirroring"). We assume a finite, open chain, and therefore couplings $J_{ij}$ with $|j-i| > N-j$ do not have an image under mirroring. 
\item (Incommensurateness) The coupling strengths $J_{ij}$ are, in general, mutually indivisible irrational numbers whose sums and differences are also irrational and mutually distinct, e.g. for $i\ne j, k\ne l$, $J_{ik}\pm J_{jk} \ne J_{il}\pm J_{jl}$ (with a very small set of exceptions due to, say, symmetry). 
\end{enumerate}

The mean $\langle E \rangle_{\beta,\gamma}$ consists of three parts corresponding to the terms $E_I, E_{II}, E_{III}$. Performing the $\beta$ integral first, we see that $\langle E_{II} \rangle_{\beta,\gamma} = 0$. Next, we may argue that in products of the form $\prodl{k}{}{\cos(2\gamma J_{ik})}$, the cosine factors are of degree one if they have no mirror images, and degree two otherwise. The only way to have a non-zero expectation is if all terms are systematically paired up by mirroring, so that the overall product is quadratic in a product of cosines. For the summand in $E_I$, this can only happen if $N$ is odd and $i$ is exactly at the center of the chain, in which case the average is $B/2^{(N-1)/2}$. When $N$ is even, the mean is 0. Finally, for general $i,j$ the last term is zero by property 2, since the cosines are generically incommensurate and therefore barring very few exceptions, most phases do not cancel out. However, in the special case that $i,j$ are mirror images, i.e. $i = N-j$, we have perfectly paired terms when $N$ is even (and one unpaired term at $k=\lfloor{N/2}\rfloor$ when $N$ is odd). Counting all occurrences of this case, the mean is approximately $\frac{1}{2^{N/2+1}}\suml{i=1}{N}{J_{i(N-i)}}\lesssim NJ_0/2^{N/2}$ where $J_0$ is the nearest-neighbor coupling in the chain. Note that asymptotically in $N$, $\langle E\rangle_{\beta,\gamma}\sim O(N/2^{N/2})$ which approaches $0$ in the infinite $N$ limit.

Next, we estimate the term $\langle E\rangle_{\beta,\gamma}^2$. By the orthogonality of trigonometric polynomials in $\beta$, we first have that $\langle E\rangle_{\beta,\gamma}^2 = \langle E_I\rangle_{\beta,\gamma}^2 + \langle E_{II}\rangle_{\beta,\gamma}^2 + \langle E_{III}\rangle_{\beta,\gamma}^2$. Therefore, we estimate each term separately. As before, we require that the cosines pair up so that their phases can cancel. First, we have 
\begin{equation}
\langle E_I\rangle_{\beta,\gamma}^2 = B^2\suml{i,j}{}{\prodl{k=1}{N}{\cos(2\gamma J_{ik})\cos(2\gamma J_{jk})}}
\end{equation}

Each summand is a product of $2N$ cosines, and only survives averaging if every cosine is paired. This happens exactly when either $i=j$ or $i = N-j$ (There is also the ``disconnected" contribution that cancels with the mean). In each case, the squared cosines give a factor of $1/2$ from averaging. Moreover, using mirror symmetry we can have fourth powers of some of the cosines, which give a factor $3/8$ from averaging. In all, the mean (minus the disconnected part) is no greater than
\begin{equation}
\langle E_I\rangle_{\beta,\gamma}^2 \lesssim 4NB^2\paren{\frac{3}{8}}^{(N-1)/2}
\end{equation}
A similar reasoning for $E_{II},E_{III}$ give us the following estimates:
\begin{align}
\langle E_{II}\rangle_{\beta,\gamma}^2 &\lesssim \frac{1}{4}NJ_0^2\paren{\frac{3}{8}}^{(N-1)/2}\\
\langle E_{III}\rangle_{\beta,\gamma}^2 &\lesssim \frac{3}{16}NJ_0^2\paren{\frac{3}{8}}^{(N-1)/2}
\end{align}
Finally, this gives
\begin{equation}
  \langle E\rangle_{\beta,\gamma}^2 \lesssim N\paren{\frac{3}{8}}^{N/2}\brac{8B^2 + J_0^2} \sim O(N\cdot \paren{3/8}^{N/2})
\end{equation}

Therefore, we see that the standard deviation $\sigma_\eta=\sigma_E/\Delta \sim \frac{\sqrt{8B^2 + J_0^2}}{\Delta}\cdot N^{1/4}\paren{3/8}^{N/4}$, which is exponentially suppressed for large $N$. For $N=20$ ions, we have $N^{1/4}\cdot\paren{3/8}^{N/4} \sim 0.02$. While this is already small, the normalization $\frac{\sqrt{8B^2 + J_0^2}}{\Delta}$ will have an additional linear $N$ factor in the denominator, making the scale for $\eta$ about $0.002$. Therefore, a typical final QAOA performance of $\eta \gtrsim 0.95$ is several standard deviations above a typical $\eta_0 \sim 0.85$. 
}

\subsection{Evidence for hardness of sampling from general QAOA circuits}
In this section we expand upon previous work \cite{Farhi2014} that gives evidence for exact sampling hardness of QAOA circuits, using the techniques of Refs.~\cite{Bremner2016,Bouland2019} to give evidence for hardness of approximate sampling.
First we relabel the bases $Y\rightarrow X \rightarrow Z$ so that the $p=1$ experiment is equivalent to preparing a state $\ket{\psi_0} = \ket{\!\!\uparrow}^{\otimes N}_x $, evolving under a Hamiltonian $H_z$ diagonal in the computational basis, followed by a uniform rotation $\tilde{H} = e^{-i\beta \sum_i \sigma^x_i}$ and measurement in the computational basis.
Following Ref.~\cite{Farhi2014}, it suffices to consider QAOA circuits with $\beta = \pi/4$.
The output state is $ \tilde{H}^{\otimes N} e^{-i\gamma H_z} H^{\otimes N} \ket{0^N}$ for some cost function $C$ diagonal in the computational basis.
\subsubsection{Generalized gap of a function}
The main idea behind proving exact sampling hardness is to examine a particular output amplitude, say the amplitude of the $\ket{0^N}$ basis state.
In Ref.~\cite{Bremner2016}, the output state after a so-called IQP circuit (which only differs from the one here in that the final rotation is a global Hadamard $H^{\otimes N}$ instead of $\tilde{H}^{\otimes N}$) has an amplitude proportional to a quantity known as the \emph{gap} of a Boolean function, $\mathrm{gap}(f) = \sum_{x:f(x)=0}1 - \sum_{x:f(x)=1}1$, the difference in the number of inputs that map to 1 and the number of inputs that map to 0 under $f$.
Finding the gap of a general function is a $\mathsf{GapP}$-complete problem.
This is a very hard problem since the class $\mathsf{GapP}$ includes $\mathsf{\#P}$, which in turn includes the whole of $\mathsf{NP}$.
The authors of Ref.~\cite{Bremner2016} prove that the gap of a degree-3 polynomial over $\mathbb{Z}_2$, $f$, may be expressed as an output amplitude of an IQP circuit.
They also show that the finding the gap of such functions $f$ is still $\mathsf{GapP}$-complete.
Following Ref.~\cite{Bremner2016}, we examine the $\ket{0^N}$ output amplitude of a QAOA state:
\begin{align}
 \bra{0^N} \tilde{H}^{\otimes N} e^{-i\gamma H_z} H^{\otimes N} \ket{0^N} =  \frac{1}{2^N}\sum_{x,y} \bra{y} i^{\sum_i y_i + \tilde{f}(x)} \ket{x},
\end{align}
where now we define the function $\tilde{f}$ to have the range $\mathbb{Z}_4$ and the Hamiltonian $H_z$ satisfies $e^{-i\gamma H_z}\ket{x} = i^{\tilde{f}(x)} \ket{x}$ for a computational basis state $\ket{x}$.
The output amplitude is thus proportional to a `generalized gap' $\mathrm{ggap}(f) := \sum_{x:f(x)=0}1 + i\sum_{x:f(x)=1}1 + i^2 \sum_{x:f(x)=2}1 + i^3\sum_{x:f(x)=3}1$ of a function $f(x) = \tilde{f}(x) + \mathrm{wt}(x) $, where $\mathrm{wt}(x)$ is the Hamming weight of $x$.
This modified function $f(x)$ is also a degree-3 polynomial over $\mathbb{Z}_4$.
Note that this restriction to degree-3 comes from the fact that the gates $Z$, $CZ$ and $CCZ$ are universal for classical computation (indeed, the Toffoli alone is universal for classical computation) and there is a natural degree-3 polynomial coming from this construction.
The quantity we have defined, $\mathrm{ggap}(f)$, can be easily shown to be $\mathsf{GapP}$-hard to compute, by reducing $\mathrm{gap}$ to $\mathrm{ggap}$.
This suffices for exact sampling hardness assuming the polynomial hierarchy ($\mathsf{PH}$) does not collapse.
\subsubsection{Approximate sampling hardness}
\renewcommand{\figurename}{\textbf{Figure}}
\renewcommand{\thefigure}{S4}
\begin{figure*}[t]
\centering
\includegraphics[width=2.0\columnwidth]{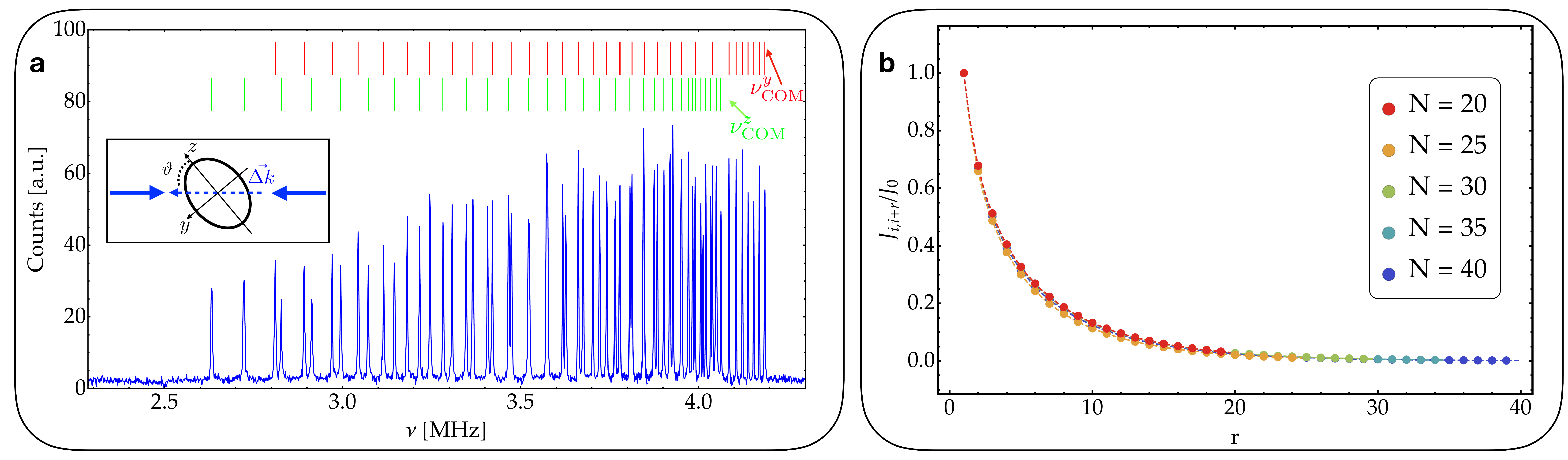}
\caption{{\bf System 2 characterization.} (a) Sideband resolved spectroscopy of a 32 ion chain with frequencies $\nu^y_{\rm COM}=4.18$~MHz and $\nu^z_{\rm COM}=4.06$~MHz, with both transverse families identified. Inset: geometrical configuration of the global Raman beams (blue arrows) with respect to the transverse principal axes of the trap (black arrows). The ellipsoid shows qualitatively an equipotential surface of the trap. (b) Average spin-spin interaction matrix element $J_{i,i+r}$ as a function of ion separation $r=|i-j|$ for the data taken in Fig. 2c in the main text, calculated with the system parameters directly measured with sideband spectroscopy, using Eq. (\ref{eq_JijCryo}). The results are normalized to the average nearest-neighbour coupling $J_0$ for each system size.}
\label{fig_normalmodes}
\end{figure*}

For approximate sampling hardness, we need two other properties, namely anti-concentration and a worst-to-average case reduction.
Anti-concentration of a circuit roughly says that the output probability is sufficiently spread out among all possible outcomes so that not many output probabilities are too small.
We choose a random family of QAOA circuits by choosing $H_z$ such that the function $f(x)$ is a degree-3 polynomial $\sum_{i,j,k} a_{i,j,k} x_ix_jx_k + \sum_{i,j} b_{i,j} x_ix_j + \sum_i c_i x_i$ with uniformly random weights $b_{i,j}$ and $c_{i} \in \mathbb{Z}_4$.
Anti-concentration then follows from the Paley-Zygmund inequality and Lemma 4 of the Supplemental Material of Ref.~\cite{Bremner2016} (with $r=s=4$).

Finally, we need to show that the problem of approximating the generalized gap is average-case hard.
Currently, no scheme for quantum computational supremacy has achieved this, and the best known result in this direction is in Ref.~\cite{Bouland2019}, where the authors show a worst-to-average case reduction for the problem of \emph{exactly} computing an output probability of a random quantum circuit.
The authors remark that their techniques may be extended to any distribution parametrized by a continuous variable.
In principle, we have such a parameter $\gamma$ available here, which continuously changes the parameters $b_{i,j}$ and $c_i$.
However, we have only shown anti-concentration when the weights $b_{i,j}$ and $c_i$ are chosen from a finite set.
It remains to be seen whether one can have the property of anti-concentration and average-case hardness holding at the same time for some specific QAOA output distribution.

\subsection{Trapped-ion experimental systems}
%
In this work two quantum simulators have been used, referred to as system 1 and 2. System 1 \cite{Kim2009} is a room-temperature ion-trap apparatus, consisting of a 3-layer linear Paul trap with transverse center-of-mass (COM) motional frequency $\nu_\text{\rm COM}=4.7$~MHz and axial center-of-mass frequencies $\nu_x$ ranging from $0.39$ to $0.6$~MHz depending on the number of trapped ions. In this system Langevin collisions with the residual background gas in the ultra high vacuum (UHV) apparatus are the main limitation to ion chain lifetime  \cite{Wineland1998}. These events can melt the crystal and eject the ions from the trap because of rf-heating or other mechanisms.

System 2 \cite{Pagano2019} is a cryogenic ion-trap apparatus based on a linear blade trap with four segmented gold coated electrodes. The trap is held at 6.5 K in a closed cycle cryostat, where differential cryo-pumping reduces the background pressure at low $10^{-12}$ Torr level, which allows for long storage times of large ion chains. For this reason system 2 has been used to perform the QAOA with a large number of qubits (Fig. 2b) or when a large number of measurements was required (Fig. 4). The two transverse trap frequencies are $\nu^y_{\rm COM}=4.4$~MHz and $\nu^z_{\rm COM}=4.26$~MHz, and the axial frequency ranges from $0.27$ to $0.46$~MHz.

\subsubsection{State preparation}

The qubit is initialized by applying resonant 369.5 nm light for about 20 $\mu$s to optically pump into the $\ket{\!\!\downarrow}_z$ state. To perform global rotations in the Bloch sphere, we apply two far-detuned, non-copropagating Raman beams whose beatnote is tuned to the hyperfine splitting $\nu_{0} = 12.642821$~GHz of the clock states $^2S_{1/2}\ket{F=0, m_F=0}$ and $^2S_{1/2}\ket{F=1,m_F=0}$ encoding the qubit \cite{Olmschenk2007}.
\noindent State preparation in our implementation of the QAOA requires qubit initialization in the $\ket{\!\!\downarrow}_z$ state by optically pumping the ions and then a global rotation into the $\ket{\!\!\uparrow}_y$ state using stimulated Raman transitions. We detect the state of each ion at the end of each experimental sequence using state-dependent fluorescence, with single site resolution.
\noindent In order to improve the accuracy of global qubit rotations, we employ a composite pulse sequence based on the dynamical decoupling BB1 scheme \cite{Brown2004}. This allows us to compensate for inhomogeneity due to the Raman beam's Gaussian profile and achieve nearly $99\%$ state preparation fidelity. The BB1 four pulse sequence is:
\begin{equation}
U_1 (\pi/2) =e^{-i\frac{\pi}{2}\sigma_i^{\theta}} e^{-i\pi\sigma_i^{3\theta}} e^{-i\frac{\pi}{2}\sigma_i^{\theta}} e^{-i\frac{\pi}{4}\sigma_i^{x}}, \nonumber \\
\end{equation}
where after the first $\pi/2$ rotation $e^{-i\frac{\pi}{4}\sigma_i^{x}}$, three additional rotations are applied: a $\pi$-pulse along an angle $\theta = \textrm{cos}^{-1}(-1/16) = 93.6\degree$, a $2\pi$-pulse along $3\theta$, and another $\pi$-pulse along $\theta$. The axes of these additional rotations are in the $x$-$y$ plane of the Bloch sphere with the specified angle referenced to the $x$-axis.

\renewcommand{\figurename}{\textbf{Figure}}
\renewcommand{\thefigure}{S5}
\begin{figure*}[t]
\centering
\includegraphics[width=1.0\textwidth]{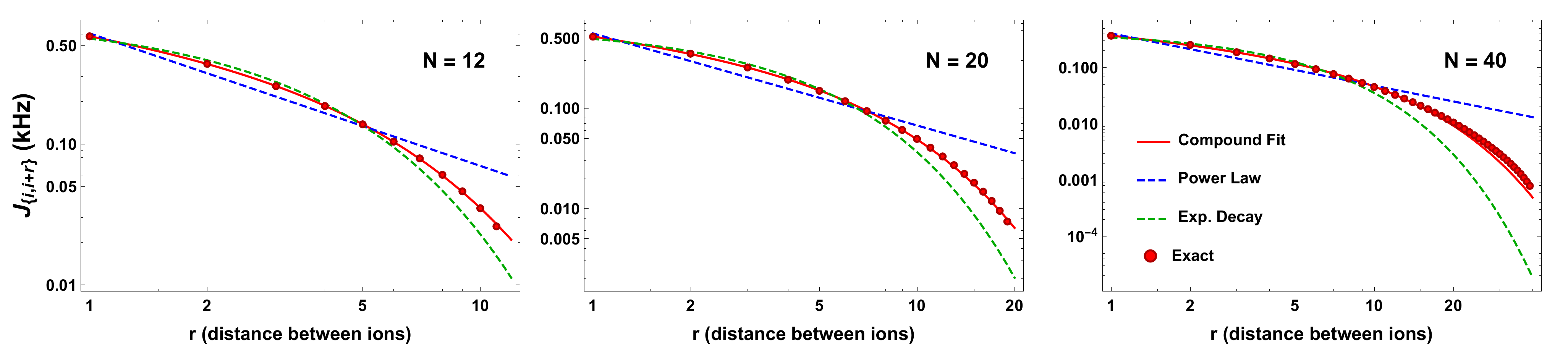}  \caption{{\bf Log-log plot of spin-spin interactions}:
red points represent the average Ising couplings between spins separated by distance $r=|i-j|$, calculated from experimental parameters using Eq. \ref{eq_FullJij}. These plots show the exact average couplings and fits corresponding to the $N=12$ and $N=20$ gradient descent experiments (Fig. 3 in the main text) and the $N=40$ exhaustive search experiment (Fig. 2c in the main text). The power law fit (blue dashed curve) fails to match the couplings for larger spin separations, as does an exponential fit (green dashed curve). The compound formula (Eq. \ref{Chimera}) fits well the actual couplings for all spin separations, even for a chain of 40 ions. The fitted parameters $\{ J_0, \alpha', \beta' \}$ for N = 12, 20, and 40 are $\{ 0.580, 0.322, 0.229 \}$, $\{ 0.517, 0.318, 0.181 \}$, and $\{ 0.369, 0.383, 0.134 \}$ respectively.
}
\label{fig:ChimeraFitCompare}
\end{figure*}

\subsubsection{Generating the Ising Hamiltonian}

We generate spin-spin interactions by employing a spin dependent force with a pair of non-copropagating 355 nm Raman beams, with a wavevector difference $\Delta k$ aligned along the transverse motional modes of the ion chain. The two off-resonant Raman beams are controlled using acousto-optic modulators which generate two interference beatnotes at frequencies $\nu_0\pm\mu$ in the M\o{}lmer-S\o{}rensen configuration \cite{Molmer1999}.
In the Lamb-Dicke regime, the laser-ion interaction gives rise to the effective spin-spin  Hamiltonian in Eq. (1) in the main text, where the coupling between the $i$-th and $j$-th ion is:
\begin{equation}
J_{ij} = \Omega^2\nu_{R} \sum_{m}\frac{b_{im}b_{jm}}{\mu^2-\nu_m^2}.
\label{eq_FullJij}
\end{equation}
Here $\Omega$ is the Rabi frequency, $\nu_R= h \Delta  k^2/(8\pi^2M)$ is the recoil frequency, $\nu_m$ is the frequency of the $m$-th normal mode, $b_{im}$ is the eigenvector matrix element for the $i$-th ion's participation to the $m$-th normal mode $(\sum_i |b_{im}|^2=\sum_m |b_{im}|^2 =1)$ \cite{James1998}, and $M$ is the mass of a single ion.

Differently from system 1, where the wavevector difference $\Delta k$ of the Raman beams is aligned along one of the principal axes of the trap, in system 2 the spin-spin interaction stems from the off-resonant coupling to both families of transverse normal modes. Eq. (\ref{eq_FullJij}) is then generalized to:
\begin{eqnarray}
J_{ij} &=& J^{y}_{ij} +J^{z}_{ij}, \nonumber\\
J^{\ell}_{ij} &=& \Omega_\ell^2\nu^{\ell}_{R} \sum_{m}\frac{b_{im}b_{jm}}{\mu^2-\left(\nu^{\ell}_m\right)^2},\,\, \ell=y,z,
\label{eq_JijCryo}
\end{eqnarray}
where $\nu^{\ell}_{R}$ is the recoil frequency given by the projection of the Raman wavevector $\Delta k$ along the two transverse principal axes of the trap $\ell=y,z$.  We infer an angle $\vartheta\sim40^o$ between $\Delta k$ and the $z$ principal axis (see inset in Fig. \ref{fig_normalmodes}a) from the ratio between the resonant spin-phonon couplings to the two transverse COM modes. Before every experiment, we perform Raman sideband cooling on both the COM and the two nearby tilt modes for both transverse mode families.

As we scale up the number of qubits (see Fig. 2c in the main text), we vary the axial confinement in order to maintain a self-similar functional form of the spin-spin interaction (see Fig. \ref{fig_normalmodes}b). For the data in Fig. 2c in the main text, we set the detuning to $\delta=\mu-\omega^y_{\rm COM}=2 \pi \times 45$ kHz and the axial frequency to $\nu_x=0.46, 0.37, 0.36, 0.31, 0.27$~MHz, for $N=20, 25, 30, 35, 40$ respectively. For the data in Fig. 4 in the main text, the detuning is $\delta/2\pi=45$ kHz and the $\nu_x=0.54$~MHz.

\subsubsection{Fitting Ising Couplings to Analytic Form}

By directly measuring trap parameters and spin-phonon couplings, we can calculate the spin-spin interaction matrix $J_{ij}$ with Eqs. (\ref{eq_FullJij}) and (\ref{eq_JijCryo}). However, in order to efficiently compute the ground state energy of the Hamiltonian in Eq. (1) (see main text) for $N\gtrsim25$ using DMRG, we approximate the Ising couplings using a translational invariant analytic function of the ion separation $r=|i-j|$. 
For $N<20$ the spin-spin coupling $J_{ij}$ between the two qubits at distance $r$ is well approximated by a power law decay:
\begin{equation}
J_{ij} \approx \frac{J_0}{r^\alpha},
\label{PowerLaw}
\end{equation}
\noindent where, as stated in the main text, $J_0$ is the average nearest-neighbor coupling and $\alpha$ is the power law exponent \cite{Porras2004}. However for larger system sizes, this approximation fails to capture the actual decay of the interaction matrix.

\noindent In order to use the DMRG algorithm to accurately compute the ground state energies, we developed a compound function to better fit our couplings. This function is a product of a power law decay and an exponential decay parametrized by $J_0$,  $\alpha'$ and $\beta'$:
\begin{equation}
J_{ij} \approx \frac{J_0}{r^{\alpha'}} e^{-\beta' (r-1)}
\label{Chimera}
\end{equation} 

\noindent As seen in Fig. \ref{fig:ChimeraFitCompare}, this functional form fits well the exact Ising couplings even for a chain of 40 ions, while both a power law and a pure exponential fit diverge significantly.
\renewcommand{\figurename}{\textbf{Figure}}
\renewcommand{\thefigure}{S6}
\begin{figure*}[t!]
\centering
\includegraphics[width=1.0\textwidth]{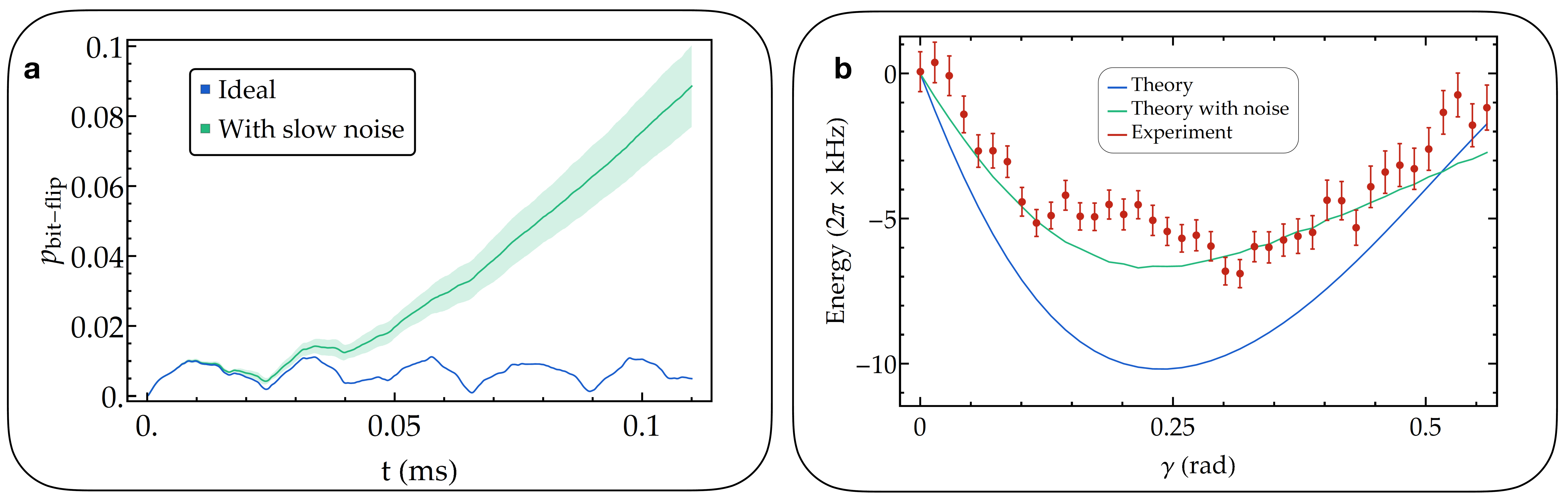}
\caption{{\bf Errors in trapped-ion quantum simulator}: (a) Phonon-assisted bit-flips per ion predicted by evolving the coherent off-resonant spin-phonon drive for 12 ions. The simulation includes slow drifts of the trap frequency and of the laser power over \blue{500} shots, each including a Hamiltonian evolution of \blue{0.11 ms, with $\delta/2\pi=45$~kHz and $\Omega/2\pi = 440$~kHz}. The shaded region is defined as the average $p_i$ plus and minus one standard deviation (see main text for details). (b) Energy as a function of the $\gamma$ parameter scan for Fig. 4 in the main text. Taking into account our total bit-flip error budget \blue{together with uncompensated light shift}, we explain most of the discrepancy between our experimental performance and the ideal QAOA energy output.}
\label{fig_bitflip}
\end{figure*}

\subsubsection{State Detection}
We detect the ion spin state by globally rotating all the spins into the measurement basis with a composite BB1 $\pi/2$ pulse as described above, to rotate the $x$ or $y$ basis into the $z$ basis), followed by the scattering of resonant laser radiation on the $^2$S$_{1/2}\ket{F=1}\leftrightarrow^2$P$_{1/2}\ket{F=0}$ cycling transition (wavelength near 369.5 nm and radiative linewidth $\gamma/2\pi\approx 20$ MHz). If the atom is projected in the $\ket{\!\!\uparrow}_z$ ``bright" state, it fluoresces strongly, while if projected in the $\ket{\!\downarrow}_z$ ``dark" state it fluoresces almost no photons because the laser is far from resonance~\cite{Olmschenk2007}.

In both systems the fluorescence of the ion chain is imaged onto an Electron Multiplying Charge Coupled Device (EMCCD) camera (Model Andor iXon Ultra 897) using an imaging objective with 0.4 numerical aperture and a magnification of 90x for both systems. The fluorescence of each ion covers roughly a 7x7 array of pixels on the EMCCD. After collecting the fluorescence for an integration time of 0.65 (1) ms for system 1 (2), we use a binary threshold to determine the state of each ion, discriminating the quantum state of each ion with near 98$\%$ (97$\%$) accuracy in system 1 (2).  The residual $2\,(3)\%$ errors include off-resonant optical pumping of the ion between spin states during detection as well as detector cross-talk between adjacent ions, readout noise, and background counts.

In system 2 the individual ion range-of-interests (ROIs) on the camera are updated with periodic diagnostic images, acquired by applying a nearly resonant cooling laser for 50 ms so that each ion fluoresces strongly regardless of its state.  The signal to background noise ratio in the diagnostic shots is larger than 100, yielding precise knowledge of the ions' center locations and  taking into account the slow $\sim 2\, \mu$m pk-pk drift due to thermal expansion/contraction of the cryostat. Ion separations range from 1.5 $\mu$m to 3.5 $\mu$m depending on the trap settings and the distance from the chain center, and are always much larger than the resolution limit of the imaging system. We utilize the pre-determined ion centers to process the individual detection shots and optimize the integration area on the EMCCD camera to collect each ion's fluorescence while minimizing cross-talk.  We estimate cross-talk to be dominated by fluorescence from nearest-neighbor, which can cause a dark ion to be erroneously read as bright.

\subsubsection{Error sources}

The fidelity of the quantum simulation is limited by experimental noise that causes the system to depart from the ideal evolution and that can have several sources that are reviewed below.
One important error source is off-resonant excitation of motional modes of the ion chain, which causes residual spin motion-entanglement. When the motion is traced out at the end of the measurement this results in a finite probability of an unwanted bit-flip. The probability of this error to occur on the $i$th ion \cite{Kim2009} is proportional to $p_i\sim\sum_{m=1}^N \left(\eta_{i m}\Omega/\delta_m \right)^2$, where $\eta_{im}=b_{im}\sqrt{\nu_R/\nu_{\rm COM}}$ (see Eq. (\ref{eq_FullJij})) and  $\delta_m=\mu-\omega_m$ is the beatnote detuning from the $m$-th normal mode. We trade off a lower error for a weaker spin-spin coupling by choosing a $\delta_{\rm COM}$ such that $\left(\eta_{\rm COM}\Omega/\delta_{\rm COM}\right)^2\lesssim1/10$. By considering the off-resonant contributions of all the modes (see Fig. \ref{fig_bitflip}), we estimate the phonon error to cause about $1\%$ bit-flip per ion. Additionally, this effect is amplified by fluctuations in the trap frequency and laser light intensity at the ions' location, increasing the probability of a phonon-assisted bit-flip event. To take this into account, we included slow drifts and fluctuations of the trap frequency and of the laser power on the timescale of \blue{500} experimental repetitions assuming noise spectral density falling as $1/f$. \blue{Given our typical trap frequency and laser power fluctuations, we assume a relative standard deviation $\Delta\Omega/\Omega\sim2 \%$ and \blue{$\Delta\delta_{\rm COM}/\delta_{\rm COM}\sim9\%$} over the timescale required to average over quantum projection noise and we end up estimating an average bit-flip probability \blue{$p_i\sim 9\%$} (see Fig. \ref{fig_bitflip}a). Moreover laser intensity, beam steering and trap frequency slow drifts over the time scale of a few hours required for data-taking cause averaging over different Ising parameters $J_0$. In particular, beam steering fluctuations create an imbalance between the red and blue $\nu_0\pm\mu$ beatnotes at the ions, producing an effective $B_z$ noisy field, that has been measured to be as high as $0.3 J_0$. To take into account these drifts, we calculated several evolutions sampling from a gaussian distribution of values of $B_z$ and $J_0$, using as a variance the standard deviations ($\sigma_{J_0}= 0.02 J_0 $ and $\sigma_{B_z}= 0.3 J_0$) observed in the experiment.} Another source of bit-flip errors is imperfect detection. Off-resonant pumping limits our average detection fidelity to $98\%($97\%$)$ for system 1 (2). A detection error is equivalent to a random bit-flip event so the two errors will sum up. A specific source of noise in system 2 is mechanical vibrations at 41 Hz and 39 Hz due to residual mechanical coupling to the cryostat \cite{Pagano2019}. This is equivalent to phase-noise on the Raman beams, which leads to dephasing of the qubits. Other less important noise sources are related to off-resonant Raman scattering errors during the Ising evolution (estimated in $7\cdot10^{-5}$ per ion) and RF heating of the transverse COM motional mode of the ion chain in system 1. 

In Fig.~\ref{fig_bitflip}b, we plot the experimentally measured energy as a function of $\gamma$, and the corresponding theoretical curves with and without incorporating errors. Using \blue{the time dependent average bit-flip probability evolution that we estimated from our error model considering phonons and detection errors and averaging over slow drifts in experimental parameters $J_0$ and $B_z$}, we get a good agreement with the experimental data (see also Fig. 2c in the main text, where the same parameters have been used), \blue{showing that we have a good understanding of the noise sources in our system.}

\end{document}